\documentclass[aps,prc,twocolumn,showpreprintnumbers,floatfix,nofootinbib]{revtex4-1}
\usepackage{amsmath,amssymb,graphicx,bm}
\usepackage{bbm}
\usepackage{color,ulem}
\usepackage{slashed}
\allowdisplaybreaks[4] 
\newcommand{\vect}[1]{\bm{\mathrm{#1}}}
\newcommand{\pv}{\vect{p}}
\newcommand{\qv}{\vect{q}}
\newcommand{\sigmav}{\vect{\sigma}}
\newcommand{\sv}{\vect{s}}
\newcommand{\nablav}{\vect{\nabla}}
\newcommand{\laplacian}{\Delta}
\newcommand{\tauv}{a}

\DeclareMathOperator{\Imag}{Im}
\renewcommand{\Im}{\Imag}
\newcommand{\ncdot}{\!\cdot\!}
\newcommand{\ntimes}{\!\!\times\!}
\renewcommand{\emph}[1]{\textit{#1}}

\begin{document}

\title{Energy and Angle Dependence of Neutrino Scattering Rates in Proto-Neutron Star and Supernova Matter within Skyrme RPA} 

\author{Mingya Duan} 
\affiliation{Universit\'e Paris-Saclay, CNRS/IN2P3, IJCLab, 91405 Orsay, France}
 
\author{Michael Urban} 
\email{michael.urban@ijclab.in2p3.fr}
\affiliation{Universit\'e Paris-Saclay, CNRS/IN2P3, IJCLab, 91405 Orsay, France}

\begin{abstract}
Supernova explosions are the most powerful neutrino sources. The neutrino emission is also the dominating cooling mechanism for a proto-neutron star, whose interior is mainly composed of extremely dense and hot nuclear matter. Neutrino transport is an essential part of the simulation of these phenomena, and modern codes are able to implement inelastic neutrino scattering and also to some extent its angle distribution. We therefore study the energy and angle dependence of neutrino scattering rates in proto-neutron star and supernova matter with the full Skyrme RPA response functions. We confirm earlier findings obtained in the Landau approximation that the RPA reduces neutrino scattering, but the detailed differential scattering rates in hot and dense matter depend sensitively on the adopted interaction. The scattering angle distribution is different for different interactions because it depends strongly on the neutron Fermi velocity. We also find that many Skyrme interactions present an unphysical feature that the Fermi velocity of neutrons exceeds the speed of light already at relatively low densities.
\end{abstract}

\maketitle

\section{Introduction}\label{sec:introduction}
Neutrinos play a crucial role in astronomical and astrophysical research. They are one of the important parts in multi-messenger astronomy, and offer a unique opportunity to explore the universe. The study of neutrino-matter interaction in neutron stars and core-collapse supernovae is not only of great significance in itself, but may also have an impact on some other relevant research topics, such as binary neutron star mergers. 

As we all know, at the end of the evolution of a star with a mass of the order of $8-30$ solar masses ($M_{\odot}$), a core-collapse supernova occurs, leaving behind a neutron star \citep{Cerda-DuranElias-Rosa}. During this collapse, the core reaches densities comparable with nuclear saturation density ($2.7 \times 10^{14}$ g/cm$^3$, corresponding to $\rho_0 =  0.16$ nucleons/fm$^3$), but then the repulsion between nucleons overcomes gravity and the infalling matter bounces back 
\citep{Janka2017a,Janka2017b}. A small sphere that has a typical mass of about $1-2M_{\odot}$ remains and forms eventually the neutron star, with a radius of about 10 km \citep{Baym1975}, while the rest is ejected. Nevertheless, the energy of this ejected matter represents only a small fraction of the initial gravitational energy of the star, while 99\% of the energy is released in the form of neutrinos \citep{BerstenMazzali} which are produced during and after the collapse by weak-interaction processes (mostly electron capture). These neutrinos are also believed to play a crucial role in explaining the explosion of the outer shells in the supernova, as Colgate and White \citep{Colgate1966} and Arnett \citep{Arnett1966} first pointed out in 1966. When the density increases to a few times $10^{11}$ g/cm$^3$ \citep{Janka2017a}, neutrinos can be scattered frequently inside the star before they escape. It means that interactions between neutrinos and matter are very important at these densities.

Hence, in the modeling of the supernova, neutrino absorption, emission, and scattering rates are essential ingredients. In the 2000s, some researchers incorporated neutrino transport into supernova simulations \citep{Rampp2000,Mezzacappa2001,Thompson2003}. The importance of adopting a high-quality neutrino transport in supernova simulations was recognized. For this reason, improving the treatment of neutrino-matter interaction has been the aim of many recent studies, e.g., using the virial approach \cite{Horowitz2006,Horowitz2017}, taking into account relativistic kinematics \citep{Roberts2017,Fischer2020a}, employing chiral effective field theory \citep{Rrapaj2015,Vidana2022,Shin2023}, highlighting the relevance of muons \citep{Bollig2017,Guo2020,Fischer2020b} and of pions \citep{Fore2020}, and so on.

Neutrino transport is crucial not only for supernova simulations but also for the evolution of proto-neutron stars. The initial internal temperature of a proto-neutron star is of the order of $10^{11}-10^{12}$ K ($10-100$ MeV) \citep{HaenselPotekhinYakovlev,DegenaarSuleimanov}. The density and temperature are so high that nuclei are completely dissolved into uniform matter of neutrons, protons, and electrons. Nevertheless, one cannot simply compute the neutrino rates by multiplying the neutrino cross sections of free nucleons with the density. The cross sections in the medium are strongly modified because the nucleon-nucleon interactions lead to strong correlations and collective effects \citep{Burrows2006}. For instance, the divergence of the neutron absorption cross section near the liquid-gas phase transition might help to understand the explosion of supernovae \citep{Margueron2004}. From a theoretical point of view, the neutrino rates can be directly related to the so-called response functions (or current-current correlation functions), computed at the relevant density, asymmetry, and temperature.

Already in earlier studies, neutrino responses were computed by using the random phase approximation (RPA) \citep{Burrows1998,Reddy1999,Horowitz2003}, which has been widely used to describe excitations in finite and infinite nuclear systems since it was first introduced to solve the collective oscillations of the electron gas \citep{Bohm1951,Pines1952,Bohm1953}. The RPA theory was described in detail in the recent review \citep{Co2023}. Recently, the improved rates for charged current interactions were incorporated into supernova simulations \citep{Oertel2020} and the proto-neutron star evolution \citep{Pascal2022}, computed using various approximations, including RPA. However, for the neutrino-nucleon scattering via the neutral current, the inelastic scattering reactions were neglected in Ref. \citep{Oertel2020} and nucleons were treated as an ideal gas in Ref. \citep{Pascal2022}. 

Indeed, neutrino-nucleon scattering is also an important process in supernova simulations and the proto-neutron star evolution. One should not ignore the interactions between nucleons in order to get a high-quality neutrino transport. In particular, neutrino scattering rates depend not only on the initial and final neutrino energies but also on the scattering angle. Until now, literature has mostly focused on the total neutrino mean free path and not insisted on the relationship between neutrino scattering rates and energy and scattering angle. Besides, the RPA responses computed by \citep{Burrows1998,Reddy1999,Horowitz2003} are in the Landau approximation instead of the full RPA. The neutrino scattering expressions given in \citep{Burrows1998,Reddy1999} are incomplete when the full RPA responses with spin-orbit interaction are adopted.

A recent review \citep{Pastore2015} discusses the full RPA response functions with Skyrme interactions. However, we think it is important to cross-check independently the response functions obtained there from quite complicated expressions. Furthermore, our work extends Ref. \citep{Pastore2015} by showing the corresponding neutrino mean-free path also for the case of asymmetric matter and discussing the angle and energy distributions.

In this work, we first summarize in Sec. \ref{sec:formalism} the derivation of the scattering rate and the computation of the RPA response functions with Skyrme interactions for the most general case of different longitudinal and transverse spin responses in asymmetric nuclear matter in order to have a self-contained presentation. In Sec. \ref{sec:RPA response functions in ANM} we compare response functions at different levels of approximation. Then we study in Sec. \ref{sec:neutrino scattering rates} the relationship between neutrino scattering rates and scattering angle. We encounter a more general problem of Skyrme interactions in the application for neutron star research, namely that the Fermi velocity exceeds the speed of light beyond some density depending on the chosen interaction. Furthermore, we will study the energy dependence of neutrino scattering rates and the evolution of the average scattering angle with neutrino energy. Finally, we summarize and conclude in Sec. \ref{sec:conclusion}. For completeness, the parameters of the used Skyrme interactions and the expressions needed for the RPA are given in the appendix.
 
\section{Formalism} 
\label{sec:formalism}
\subsection{Neutrino-nucleon scattering in hot nuclear matter} 
\label{subsec:neutrino-nucleon scattering in hot nuclear matter}
As mentioned above, the neutrino rates can be related to the response functions. In this section, we derive the relation between the neutrino scattering rate and the nuclear density and spin-density correlation functions using Fermi's golden rule. Throughout this paper, we use units with $\hbar = c = k_B = 1$, where $\hbar$, $c$, and $k_B$ are the reduced Planck constant, the speed of light, and the Boltzmann constant, respectively.

Neutrinos scatter off nucleons via the weak neutral current. The interaction Lagrangian density is \citep{Iwamoto1982} 
\begin{equation}\label{eq:lagrangian density}
\mathcal{L}_{I}(x) = \frac{G}{\sqrt{2}} l_{\mu} (x) j^{\mu} (x),
\end{equation}
($\mu = 0, 1, 2, 3$), where $G$ is the Fermi weak interaction constant ($G \approx 8.975 \times 10^{-5}$ MeV fm$^3$), $l_{\mu}$ is the lepton weak neutral current, and $j^{\mu}$ is the weak current of the nucleons. 
The lepton current has the form
\begin{equation}\label{eq:lepton weak neutral current}
l_{\mu} (x) =\hat{\bar{\psi}}_{\nu} \gamma_{\mu} (1-\gamma_5) \hat{\psi}_{\nu},
\end{equation}
where $\hat{\psi}_{\nu}$ represents the neutrino field operator, and $\hat{\bar{\psi}} = \hat{\psi}^\dagger \gamma^0$. $\gamma_{\mu}$ are the usual $\gamma$ matrices and $\gamma^{5} = i \gamma^0 \gamma^{1} \gamma^2 \gamma^3 =\gamma_5$ \citep{BjorkenDrell}.

The neutrino field operator can be expressed as (we don't write the antineutrino part):
\begin{equation}\label{eq:neutrino field operator}
\hat{\psi}_{\nu} =\sum_{\pv}  \sqrt{\frac{m_{\nu}}{p^0 V}} u^{}_{L}( \pv) e^{-i p \cdot x} b_{\nu} (\pv),
\end{equation}
where $V$ is the volume of the system, $m_{\nu}$ is the neutrino mass (as in \cite{BjorkenDrell} we use the normalization with finite $m_\nu$ and take the limit $m_{\nu} \rightarrow 0$ in the end), $u^{}_{L}$ is the bispinor ($u^{}_{L} \overline{u}^{}_{L}=\frac{1-\gamma_5}{2} \frac{\slashed{p}{}+m^{}_{\nu}}{2m_{\nu}}$ with  $\slashed{p}{} = \gamma^{\alpha} p_{\alpha}$, the subscript $L$ indicates that neutrinos are left-handed), $p = (p^0, \pv)$ is the neutrino four-momentum with $p^0 = E_\nu(\pv) = |\pv|$, and $b_\nu$ is the neutrino annihilation operator.

The nucleon weak current is 
\begin{align}\label{eq:nucleon weak current}
j^{\mu} (x) & = \sum_{\tauv=n,p} \frac{1}{2} \hat{\bar{\psi}}^{}_{\tauv} \gamma^{\mu} (C^{}_{V,\tauv} - C^{}_{A,\tauv} \gamma^{5}) \hat{\psi}^{}_{\tauv}
\end{align}
where $\hat{\psi}_{n,p}$ are the neutron and proton field operators, and $C_{V,\tauv}$ and $C_{A,\tauv}$ are the relevant neutral current vector and axial-vector coupling constants. The values of these coupling constants are $C_{V,n} = -1$, $C_{A,n} = -1.23$ for neutrons and $C_{V,p} = 0.08$, $C_{A,p} = 1.23$ for protons \citep{Reddy1998}. 

The nucleon field operators are:
\begin{equation}\label{eq:nucleon field operators}
\hat{\psi}_{a} = \sum_{\pv s} c_{as}(\pv) \frac{1}{\sqrt{V}} \chi_{s} e^{-i p \cdot x},
\end{equation}
where $c_{as}(\pv)$ is the annihilation operator of a nucleon of type $a$ and spin $s$. We treat the nucleons as non-relativistic, i.e., $\chi_s$ is a two-component Pauli spinor corresponding to the two large components of the Dirac bispinor, and $p^0 = m_a+\tfrac{\pv^2}{2m_a}$.
Thus $j^{\mu} (x)$ can be written as \citep{Iwamoto1982}
\begin{align}\label{eq:nucleon weak current-approx}
j^{\mu} (x) \approx \sum_{\tauv=n,p} \frac{1}{2} (C^{}_{V,\tauv} \hat{\psi}_{\tauv}^{\dagger} \hat{\psi}^{}_{\tauv}, - C^{}_{A,\tauv} \hat{\psi}_{\tauv}^{\dagger} \sigmav \hat{\psi}^{}_{\tauv} )^{\mu}.
\end{align}

The interaction Hamiltonian can be given by using 
\begin{equation}\label{eq:interaction hamiltonian-scattering at finite}
\hat{H}_{I}  = \int d^3 x \mathcal{H}_{I} = - \int d^3 x \mathcal{L}_{I}.
\end{equation}
Then insert it into
\begin{equation}\label{eq:Hfi}
H_{fi}= \langle \pv'_{\nu} \lambda' \vert \hat{H}_{I} \vert \pv_{\nu} \lambda \rangle,
\end{equation}
to obtain $H_{fi}$, where $\pv_\nu, \pv_\nu', \lambda$ and $\lambda'$ denote the initial and final neutrino momenta and quantum numbers characterizing the nuclear matter, respectively.
According to Fermi's golden rule \citep{SakuraiNapolitano}, the transition rate is given by
\begin{equation}\label{eq:fermi's golden rule}
R_{\pv_{\nu}, \lambda \rightarrow \pv'_{\nu}, \lambda'}  = 2\pi \vert H_{fi} \vert^2 \delta( \omega -E_{\lambda' \lambda}), 
\end{equation}
where $\omega = E_{\nu} - E'_{\nu}$ is the energy transfer ($E_{\nu}$ and $E'_{\nu}$ are the initial and final neutrino energies), and $E_{\lambda' \lambda}$ is the difference between the final and the initial energy of the nuclear matter,  i.e., $E_{\lambda' \lambda} = E_{\lambda'} - E_{\lambda}$. Taking a statistical average over initial nucleon states and summing over final nucleon states, we can calculate the neutrino-nucleon scattering rate using the following expression:
\begin{widetext}

\begin{multline}\label{eq:scattering rate-initial state to final state}
R_{\pv_{\nu} \rightarrow \pv'_{\nu}} = \sum_{\lambda \lambda'} A_{\lambda} R_{\pv_{\nu}, \lambda \rightarrow \pv'_{\nu}, \lambda'} = \frac{G^2 \pi}{2V} \sum_{aa'=n,p} \Big\{ (1+ \cos \theta) C^{}_{V,a}C^{}_{V,a'} S_{aa'}^{(S=0)} (\qv,\omega) \\
+ [\hat{p}'_{\nu i} \hat{p}^{}_{\nu j} + \hat{p}'_{\nu j} \hat{p}^{}_{\nu i} + (1-\cos\theta) \delta^{}_{ij} + i \varepsilon^{}_{ijk} (\hat{p}^{}_{\nu k} - \hat{p}'_{\nu k})] C^{}_{A,a} C^{}_{A,a'} S_{aa',ij}^{(S=1)}(\qv,\omega)\Big\},
\end{multline}
where $A_{\lambda} = \frac{1}{Z} e^{-E_{\lambda}/T}$ ($Z$ is the partition function), summation over repeated indices is implied ($i,j,k = 1\dots 3$), $\hat{\pv}_\nu$ and $\hat{\pv}'_\nu$ are unit vectors in direction of $\pv_\nu$ and $\pv'_\nu$, respectively, $\qv = \pv_{\nu} - \pv'_{\nu}$ is the momentum transfer and $\theta$ is the scattering angle. The dynamical structure factors are defined as
\begin{gather}
\label{eq:structure factor S-s0}
S_{\tauv \tauv'}^{(S=0)} (\qv ,\omega) = \frac{1}{V} \sum_{\lambda \lambda'} A_{\lambda} \int_{V} d^3 x' e^{-i \qv \cdot \textbf{x}'} \int_{V} d^3 x e^{i \qv \cdot \textbf{x}} \langle \lambda \vert   \hat{\psi}^{\dagger}_{\tauv'} (\textbf{x}') \hat{\psi}_{\tauv'}(\textbf{x}')  \vert  \lambda' \rangle \langle \lambda' \vert   \hat{\psi}^{\dagger}_{\tauv} (\textbf{x}) \hat{\psi}_{\tauv}(\textbf{x})  \vert \lambda \rangle \delta(\omega -E_{\lambda' \lambda})\,,
\\
\label{eq:structure factor S-s1}
S_{\tauv \tauv',ij}^{(S=1)} (\qv ,\omega) = \frac{1}{V} \sum_{\lambda \lambda'} A_{\lambda} \int_{V} d^3 x' e^{-i \qv \cdot \textbf{x}'} \int_{V} d^3 x e^{i \qv \cdot \textbf{x}} \langle \lambda \vert   \hat{\psi}^{\dagger}_{\tauv'} (\textbf{x}') \sigma_{j} \hat{\psi}_{\tauv'}(\textbf{x}')  \vert  \lambda' \rangle \langle \lambda' \vert   \hat{\psi}^{\dagger}_{\tauv} (\textbf{x}) \sigma_{i} \hat{\psi}_{\tauv}(\textbf{x})  \vert \lambda \rangle \delta(\omega -E_{\lambda' \lambda})\,,
\end{gather}
with $\sigmav$ the Pauli matrices. Notice that they satisfy
\begin{equation}\label{eq:S-real}
S_{\tauv \tauv'}^{(S=0)} (\qv, \omega) = S_{\tauv' \tauv}^{(S=0)} (\qv, \omega)\quad\text{and}\quad 
S_{\tauv \tauv',ij}^{(S=1)} (\qv, \omega) = S_{\tauv' \tauv,ji}^{(S=1)} (\qv, \omega)\,,
\end{equation}
and that they are real as can be shown using time-reversal symmetry. Hence, the $\varepsilon_{ijk}$ term in Eq (\ref{eq:scattering rate-initial state to final state}) does not contribute. The volume $V$ drops out when one computes 
\begin{equation}\label{eq:scattering rate-integral}
R = \sum_{\pv'_{\nu}} R_{\pv_{\nu} \rightarrow \pv'_{\nu}}  = \int d^3 p'_{\nu} \frac{V}{(2 \pi)^3} R_{\pv_{\nu} \rightarrow \pv'_{\nu}} = \int d^3 p'_{\nu} \frac{d^3 R}{d^3 p'_{\nu}}. 
\end{equation}
The double differential scattering rate can be written as 
\begin{multline}
\label{eq:double differential of scattering rate}
\frac{d^2 R}{d \cos\theta\, d p'_{\nu}} = p_{\nu}^{\prime\,2} \frac{V}{(2 \pi)^2} R_{\pv_{\nu} \rightarrow \pv'_{\nu}}
 = \frac{G^2}{8 \pi}  p_{\nu}^{\prime\,2} \sum_{aa'=n,p} \Big\{ (1+ \cos\theta) C_{V,a}C_{V,a'} S_{aa'}^{(S=0)} (\qv,\omega)\\
 + [\hat{p}'_{\nu i} \hat{p}_{\nu j} + \hat{p}'_{\nu j} \hat{p}_{\nu i} + (1-\cos\theta) \delta_{ij}]  C_{A,a} C_{A,a'} S_{aa',ij}^{(S=1)} (\qv,\omega)\Big\}.
\end{multline}

\end{widetext}
Using $E_{\nu}=p_{\nu}$ and $E'_{\nu} = p'_{\nu}$, we can express the double differential scattering rate in terms of the energy and momentum transfers using the following relations
\begin{gather}\label{eq:relate-double differential}
\frac{d^2 R}{d q\,  d \omega} = \frac{q}{p_{\nu} p'_{\nu}} \frac{d^2 R}{d \cos\theta\, d p'_{\nu}} = \frac{q}{E_{\nu} E'_{\nu}}  \frac{d^2 R}{d \cos\theta \,d E'_{\nu}}\,,
\\
E'_{\nu} = E_{\nu} - \omega\,, \quad \cos\theta = \frac{E^2_{\nu} + E^{\prime\,2}_{\nu} - q^2}{2 E_{\nu} E'_{\nu}}\,.
\end{gather}

On the other hand, introducing the notation $\sigma_0=1$, the dynamical structure factors can be written in the following form
\begin{equation}\label{eq:structure factors and correlation functions}
S_{\tauv \tauv',ij}(\qv, \omega) = - \frac{1}{\pi} \frac{1}{1-e^{-\omega/T}} \Im \Pi_{\tauv \tauv',ij}^{R} (\vect{q}, \omega),
\end{equation}
(with $i=j=0$ for $S=0$ and $i,j=1,2,3$ for $S=1$), where $\Pi_{\tauv \tauv',ij}^{R} (\vect{q}, \omega)$ is the Fourier transform of the retarded correlation function which is defined as
\begin{multline}\label{eq:definition of correlation functions}
\Pi_{\tauv \tauv',ij}^{R} (\textbf{x},t)= -i \theta(t) \sum_{\lambda} A_{\lambda} \\
\times\langle \lambda \vert [ \hat{\psi}_{\tauv'}^{\dagger} (\textbf{x},t) \sigma_{j}  \hat{\psi}_{\tauv'} (\textbf{x},t), \hat{\psi}_{\tauv}^{\dagger} (\textbf{0},0) \sigma_{i} \hat{\psi}_{\tauv} (\textbf{0},0)] \vert \lambda \rangle\,.
\end{multline}
Thus we can relate the neutrino scattering rate to the nuclear density and spin-density correlation functions.

\subsection{Skyrme RPA response functions in asymmetric nuclear matter} 
\label{subsec:response functions in ANM}

The response functions depend on the nuclear interaction. Here we will use Skyrme interactions, because the solution of the full RPA is relatively easy in this case. Skyrme interactions, which started with the original work of Skyrme \citep{Skyrme1956}, are well-known and popular effective nucleon-nucleon interactions. In the early stage, they were mostly used in nuclear structure calculations \citep{Vautherin1972,Beiner1975,Kohler1976}. They are easy to use because they are zero-range interactions. The Skyrme energy-density functional is also widely used to describe nuclear matter. With the notations of Ref. \citep{Bender2003}, the energy-density functional is written as
\begin{align}
\varepsilon_{s} = 
& C_0^{\rho} \rho^2
+ C_1^{\rho} (\rho_n - \rho_p)^2 
+ C_0^{\tau} (\rho \tau -  \vect{j}^2)
\nonumber\\ &
 + C_1^{\tau} [(\rho_n - \rho_p) (\tau_n - \tau_p) - (\vect{j}_n - \vect{j}_p)^2] 
\nonumber\\ &
+ C_0^{\laplacian \rho} \rho \laplacian \rho
+ C_1^{\laplacian \rho} (\rho_n -\rho_p) \laplacian (\rho_n - \rho_p)
\nonumber\\ &
+ C_0^{\nabla\rho} (\nablav\rho)^2
+ C_1^{\nabla\rho} [\nablav(\rho_n-\rho_p)]^2 
\nonumber\\
&+ C_0^{s} \sv^2 
+ C_1^{s} (\sv_n - \sv_p)^2  
+ C_0^{sT} ( \sv \cdot \vect{T} -\mathbbm{J}^2) 
\nonumber\\ &
+ C_1^{sT} [ (\sv_n - \sv_p) \cdot (\vect{T}_n - \vect{T}_p) - (\mathbbm{J}_n - \mathbbm{J}_p)^2] 
\nonumber\\ &
+ C_0^{\laplacian s} \sv \cdot \laplacian \sv 
+ C_1^{\laplacian s} (\sv_n - \sv_p) \cdot \laplacian (\sv_n - \sv_p)   
\nonumber\\&
+ C_0^{\nabla\otimes s} (\nablav\otimes \sv)^2
+ C_1^{\nabla\otimes s} [\nablav\otimes(\sv_n-\sv_p)]^2 
\nonumber\\ &
+ C_0^{\nabla s} (\nablav\cdot\sv)^2
+ C_1^{\nabla s} [\nablav\cdot(\sv_n-\sv_p)]^2
\nonumber\\ &
+ C_0^{\nabla J} ( \rho \nablav \cdot \vect{J} + \sv \cdot \nablav \times \vect{j}) 
\nonumber\\ &
+ C_1^{\nabla J} [(\rho_n - \rho_p) \nablav \cdot (\vect{J}_n - \vect{J}_p)
\nonumber\\ &
\phantom{+ C_1^{\nabla J}} + (\sv_n - \sv_p) \cdot \nablav \!\times\! (\vect{j}_n - \vect{j}_p)]\,.
\end{align}
Here we have included additional $(\nablav \rho)^2$ and $(\nablav\otimes\sv)^2 \equiv \sum_{ij} (\nabla_i s_j)^2$ terms that are absent in Ref. \cite{Bender2003} but that are needed in the case of generalized Skyrme functionals.
Then the energy functional is given by integration
\begin{equation}\label{eq:energy functional}
E_{s} = \int d^3 r \varepsilon_{s}.
\end{equation}

In the following description, let us follow the notation in Ref. \citep{Urban2020}. The residual particle-hole (ph) interaction is derived by computing 
\begin{equation}\label{eq:ph-interaction}
  \mathcal{V}_{21}^0 = \frac{\delta ^2 E_{s}}{\delta \rho^{}_{2' 2} \delta \rho^{}_{1 1'}},
\end{equation}
with the short-hand notation
\begin{align}\label{eq:notation1}
1 =\Big(\tauv^{}_1, \pv^{}_1+\frac{\qv}{2},s^{}_1\Big),&& 1' =\Big(\tauv^{}_1, \pv^{}_1-\frac{\qv}{2},s'_1\Big), 
\end{align}
and analogous for $2$ and $2'$.
$\rho^{}_{1 1'} = \langle c_{1'}^{\dagger} c^{}_1 \rangle$ denotes the density matrix. Using this procedure, we automatically include direct and exchange terms of the ph interaction \citep{Garcia-Recio1992}. There is a minor difference between the case of asymmetric nuclear matter and that of pure neutron matter. The general form of the ph interaction in asymmetric nuclear matter is 
\begin{align}\label{eq:ph interaction-ANM}
\mathcal{V}_{21}^0 = & v_1^0 (q) + v_2^0 p_1^2 +v_{28}^0  p_2^2 + v_3^0 \pv_1 \cdot \pv_2 \nonumber \\
& +[ v_4^0 (q) + v_5^0 (p_1^2 + p_2^2 ) + v_6^0 \pv_1 \cdot \pv_2 ] \sigmav_1 \cdot \sigmav_2 \nonumber \\
& + v_8^0 i \qv \cdot (\pv_1 - \pv_2) \times (\sigmav_1 + \sigmav_2).
\end{align}
(The reason for the unusual numbering of the coefficient $v_{28}^0$ is that we want to keep the same labeling for the other terms as in \citep{Urban2020}.) In general, $v_{28}^0$ is no longer equal to $v_2^0$. Actually, $v_{28}^0 = v_2^0 $ is still true except that $v_{28,np}^0 = v_{2,pn}^0 \neq v_{2,np}^0 = v_{28,pn}^0$ for some Skyrme interactions like BSk19, BSk20, and so on.

To obtain the RPA vertex $\mathcal{V}$, we need to solve the Bethe-Salpeter-like equation:
\begin{equation}\label{eq:BS-like equation}
\mathcal{V}_{21} = \mathcal{V}_{21}^0 - \sum_3 \mathcal{V}_{23}^0 G_{\rm ph} (\pv_3, \qv) \mathcal{V}_{31},
\end{equation}
where $\sum_3 = \sum_{s^{}_3s'_3 \tauv^{}_3} \int d^3 p^{}_3/(2\pi)^3$, the particle-hole Green's function is defined as
\begin{equation}\label{eq:ph green's function}
G_{\rm ph}^{\tauv} (\pv,\qv, \omega) = \frac{n_{ \pv + \frac{\qv}{2}}^{\tauv} - n_{ \pv -\frac{\qv}{2}}^{\tauv}}{\omega - (  \epsilon_{ \pv +\frac{\qv}{2}}^{\tauv} - \epsilon_{ \pv -\frac{\qv}{2}}^{\tauv}  ) +i \eta},
\end{equation}
where $n^a_{\pv} = 1/(e^{(\epsilon^a_{\pv}-\mu_a)/T}+1)$
denotes the finite-temperature occupation number and $\epsilon^a_{\pv} = \frac{p^2}{2m^*_a}+U_a$ the HF single-particle energy, with $\mu_a$ the chemical potential, $U_a$ the mean field, and $m^*_a$ the effective mass of nucleons of kind $a$.

The number of terms will increase compared to Ref. \citep{Urban2020} for two reasons. One is because the energy transfer $\omega$ is not zero, and the other is because the number of independent terms cannot be reduced as in the pure neutron matter case. The RPA vertex is finally determined as
\begin{align}\label{eq:RPA vertex-ANM}
\mathcal{V}_{21} = & v_1 +v_2 p_1^2  + v_3 \pv_1 \ncdot \pv_2 + v_4 \sigmav_1 \ncdot \sigmav_2  + v_5 \sigmav_1 \ncdot \sigmav_2 p_1^2 \nonumber\\
& + v_6 \sigmav_1 \ncdot \sigmav_2 \pv_1 \ncdot \pv_2  +v_7 \sigmav_1 \ncdot \qv \sigmav_2 \ncdot \qv  + v_8 i \qv \ncdot \pv_1 \ntimes \sigmav_1  \nonumber\\
& + v_9 i \qv \ncdot \pv_1 \ntimes \sigmav_2  + v_{10} p_1^2 p_2^2  + v_{11} \pv_1 \ncdot \qv \pv_2 \ncdot \qv \nonumber\\
& + v_{12} \sigmav_1 \ncdot \sigmav_2 p_1^2 p_2^2  + v_{13} \sigmav_1 \ncdot \sigmav_2 \pv_1 \ncdot \qv \pv_2 \ncdot \qv \nonumber\\
& + v_{14} \sigmav_1 \ncdot \qv \sigmav_2 \ncdot \qv p_1^2  +v_{15} \sigmav_1 \ncdot \qv \sigmav_2 \ncdot \qv p_1^2 p_2^2  \nonumber\\
& - v_{16} i \qv \ncdot  \pv_2 \ntimes \sigmav_2 p_1^2  - v_{17} i \qv \ncdot  \pv_2 \ntimes \sigmav_1 p_1^2 \nonumber\\
& + v_{18} \qv \ncdot \pv_1 \ntimes \sigmav_1 \qv \ncdot \pv_2 \ntimes \sigmav_2 + v_{19} \pv_1 \ncdot \qv \nonumber\\
& + v_{20} \pv_1 \ncdot \qv p_2^2  + v_{21} \sigmav_1 \ncdot \sigmav_2  \pv_1 \ncdot \qv  + v_{22}  \sigmav_1 \ncdot \sigmav_2  \pv_1 \ncdot \qv p_2^2  \nonumber\\
& + v_{23} \sigmav_1 \ncdot \qv \sigmav_2 \ncdot \qv \pv_1 \ncdot \qv  + v_{24}  \sigmav_1 \ncdot \qv \sigmav_2 \ncdot \qv \pv_1 \ncdot \qv p_2^2 \nonumber\\
& + v_{25} \sigmav_1 \ncdot \qv \sigmav_2 \ncdot \qv \pv_1 \ncdot \qv \pv_2 \ncdot \qv  - v_{26} i \qv \ncdot \pv_2 \ntimes \sigmav_2  \pv_1 \ncdot \qv \nonumber\\
& - v_{27} i \qv \ncdot  \pv_2 \ntimes \sigmav_1 \pv_1 \ncdot \qv  +v_{28} p_2^2  + v_{29} \sigmav_1 \ncdot \sigmav_2 p_2^2 \nonumber\\
& -v_{30} i \qv \ncdot \pv_2 \ntimes \sigmav_2 - v_{31} i \qv \ncdot \pv_2 \ntimes \sigmav_1 + v_{32} \sigmav_1 \ncdot \qv \sigmav_2 \ncdot \qv p_2^2 \nonumber\\
& + v_{33} i \qv \ncdot  \pv_1 \ntimes \sigmav_1 p_2^2 +  v_{34} i \qv \ncdot  \pv_1 \ntimes \sigmav_2 p_2^2  + v_{35} \pv_2 \ncdot \qv \nonumber\\
& + v_{36} \pv_2 \ncdot \qv p_1^2 + v_{37} \sigmav_1 \ncdot \sigmav_2  \pv_2 \ncdot \qv + v_{38}  \sigmav_1 \ncdot \sigmav_2  \pv_2 \ncdot \qv p_1^2 \nonumber\\
& + v_{39} \sigmav_1 \ncdot \qv \sigmav_2 \ncdot \qv \pv_2 \ncdot \qv + v_{40}  \sigmav_1 \ncdot \qv \sigmav_2 \ncdot \qv \pv_2 \ncdot \qv p_1^2 \nonumber\\
& +v_{41} i \qv \ncdot \pv_1 \ntimes \sigmav_1  \pv_2 \ncdot \qv + v_{42} i \qv \ncdot  \pv_1 \ntimes \sigmav_2 \pv_2 \ncdot \qv.
\end{align}
Following the steps explained in \citep{Urban2020}, we define the generalized Lindhard functions
\begin{align}\label{eq:lindhard functions}
& \Pi_{k}^{\tauv}(q, \omega) = - 2 \int \frac{d^3 p}{(2\pi)^3} p^{k} G_{\rm ph}^{\tauv} (\pv, \qv, \omega), \\
& \Pi_{2L}^{\tauv} (q, \omega) = - 2 \int \frac{d^3 p}{(2 \pi)^3} p^2 \cos^2 \vartheta \, G_{\rm ph}^{\tauv} (\pv, \qv, \omega), \\
& \Pi_{2T}^{\tauv} (q, \omega)  = \frac{\Pi_2^{\tauv} - \Pi_{2L}^{\tauv}}{2},
\end{align}
where $\vartheta$ is the angle between $\pv$ and $\qv$, to compute the coefficients $v_i$. Besides, a useful relation can be shown:
\begin{equation}\label{eq:useful computation}
- 2 \int \frac{d^3 p}{(2\pi)^3} \pv p^{k} G_{\rm ph}^{\tauv} (\pv, \qv, \omega) =\qv \frac{m_{\tauv}^* \omega}{q^2} \Pi_{k}^{\tauv} (q, \omega).
\end{equation}
To obtain $v_i$, we solve the following linear system of equations
\begin{multline}
    \sum_k \begin{pmatrix}
    \delta_{ik}-A_{ik,nn}&-A_{ik,np}\\
    -A_{ik,pn}&\delta_{ik}-A_{ik,pp}
    \end{pmatrix}
    \begin{pmatrix}
    v_{k,nn}&v_{k,np}\\v_{k,pn}&v_{k,pp}
    \end{pmatrix}\\
    = \begin{pmatrix}
    v_{i,nn}^0&v_{i,np}^0\\v_{i,pn}^0&v_{i,pp}^0
    \end{pmatrix},
\end{multline}
where the matrix elements $A_{ik,\tauv \tauv'}$ for the different $\tauv \tauv'$ combinations have the same form, only the effective mass $m^*$, $v_i^0$, and $\Pi_i$ are different. The expressions for the $v_{i,\tauv \tauv'}^0$ and $A_{ik}$ are given in the appendix.

The full RPA response functions in asymmetric nuclear matter can be given after finishing the numerical computation of the coefficients $v_i$ by using the definition
\begin{multline}\label{eq:Pi-s1}
\Pi_{{\rm RPA},\tauv \tauv',ij}= -\sum_1 \sigma_{1i} G_{\rm ph}^{\tauv'} (\pv_1,\qv) \sigma_{1j} \\
+ \sum_{1,2} \sigma_{2j} G_{\rm ph}^{\tauv}(\pv_2,\qv) \mathcal{V}_{21,\tauv \tauv'} G_{\rm ph}^{\tauv'} (\pv_1,\qv) \sigma_{1i},
\end{multline}
where as in Eq. (\ref{eq:structure factors and correlation functions}) $i=j=0$ for $S=0$ and $i,j = 1,2,3$ for $S=1$. The density response functions ($S=0$) are
\begin{align}\label{eq:Pi-s0}
&\Pi_{{\rm RPA},\tauv \tauv'}^{(S=0)} 
\nonumber\\
&\mbox{}
\nonumber\\
&\quad = \Pi_0^{\tauv'} \delta_{\tauv \tauv'} 
+ v_{1,\tauv \tauv'} \Pi_0^{\tauv'} \Pi_0^{\tauv} 
+ v_{2,\tauv \tauv'} \Pi_2^{\tauv'} \Pi_0^{\tauv} 
\nonumber\\
&\qquad + \frac{m_{\tauv'}^* m_{\tauv}^* \omega^2}{q^2} v_{3,\tauv \tauv'} \Pi_0^{\tauv'} \Pi_0^{\tauv} 
+ v_{10,\tauv \tauv'} \Pi_2^{\tauv'} \Pi_2^{\tauv} 
\nonumber\\
&\qquad +  m_{\tauv'}^* m_{\tauv}^* \omega^2 v_{11,\tauv \tauv'} \Pi_0^{\tauv'} \Pi_0^{\tauv} 
+ m_{\tauv'}^* \omega v_{19,\tauv \tauv'} \Pi_0^{\tauv'} \Pi_0^{\tauv} 
\nonumber\\
&\qquad + m_{\tauv'}^* \omega v_{20,\tauv \tauv'} \Pi_0^{\tauv'} \Pi_2^{\tauv} 
+ v_{28,\tauv \tauv'} \Pi_0^{\tauv'} \Pi_2^{\tauv}
\nonumber\\
&\qquad + m_{\tauv}^* \omega v_{35,\tauv \tauv'} \Pi_0^{\tauv'} \Pi_0^{\tauv}
+ m_{\tauv}^* \omega v_{36,\tauv \tauv'} \Pi_2^{\tauv'} \Pi_0^{\tauv}. 
\end{align}
The transverse spin response functions ($S=1,M=\pm1$) are
\begin{align}\label{eq:Pi-s1-m1}
&\Pi_{{\rm RPA},\tauv \tauv'}^{(S=1,M=\pm1)} 
  = \frac{1}{2} \sum_{ij=1}^3 \Pi_{{\rm RPA}, \tauv \tauv', ij}^{(S=1)} \Big(\delta_{ij} - \frac{q_i q_j}{q^2}\Big) 
\nonumber\\
&\quad = \Pi_0^{\tauv'} \delta_{\tauv \tauv'} 
+ v_{4,\tauv \tauv'} \Pi_0^{\tauv'} \Pi_0^{\tauv} 
+ v_{5,\tauv \tauv'} \Pi_2^{\tauv'} \Pi_0^{\tauv}
\nonumber\\
&\qquad + \frac{m_{\tauv'}^* m_{\tauv}^* \omega^2}{q^2} v_{6,\tauv \tauv'} \Pi_0^{\tauv'} \Pi_0^{\tauv} 
+ v_{12,\tauv \tauv'} \Pi_2^{\tauv'} \Pi_2^{\tauv}
\nonumber\\
&\qquad + m_{\tauv'}^* m_{\tauv}^* \omega^2 v_{13,\tauv \tauv'} \Pi_0^{\tauv'} \Pi_0^{\tauv} 
+ m_{\tauv'}^* \omega v_{21,\tauv \tauv'} \Pi_0^{\tauv'} \Pi_0^{\tauv}
\nonumber\\
&\qquad + m_{\tauv'}^* \omega v_{22,\tauv \tauv'} \Pi_0^{\tauv'} \Pi_2^{\tauv}
+ v_{29,\tauv \tauv'} \Pi_0^{\tauv'} \Pi_2^{\tauv}
\nonumber\\
&\qquad + m_{\tauv}^* \omega v_{37,\tauv \tauv'} \Pi_0^{\tauv'} \Pi_0^{\tauv} 
+ m_{\tauv}^* \omega v_{38,\tauv \tauv'} \Pi_2^{\tauv'} \Pi_0^{\tauv}.
\end{align}
And the longitudinal spin response functions ($S=1,M=0$) are
\begin{align}\label{eq:Pi-s1-m0}
&\Pi_{{\rm RPA},\tauv \tauv'}^{(S=1,M=0)}
= \sum_{ij=1}^3 \Pi_{{\rm RPA}, \tauv \tauv',ij}^{(S=1)} \frac{q_i q_j}{q^2}
\nonumber\\
&\quad = \Pi_{{\rm RPA},\tauv \tauv'}^{(S=1,M=\pm 1)} 
+ q^2 \Big(v_{7,\tauv \tauv'} \Pi_0^{\tauv'} \Pi_0^{\tauv}
\nonumber\\
&\qquad + v_{14,\tauv \tauv'} \Pi_2^{\tauv'} \Pi_0^{\tauv}
+ v_{15,\tauv \tauv'} \Pi_2^{\tauv'} \Pi_2^{\tauv} 
\nonumber\\
&\qquad + m_{\tauv'}^* \omega v_{23,\tauv \tauv'} \Pi_0^{\tauv'} \Pi_0^{\tauv}
+ m_{\tauv'}^* \omega v_{24,\tauv \tauv'} \Pi_0^{\tauv'} \Pi_2^{\tauv} 
\nonumber\\
&\qquad + m_{\tauv'}^* m_{\tauv}^* \omega^2 v_{25,\tauv \tauv'}\Pi_0^{\tauv'} \Pi_0^{\tauv} 
+ v_{32,\tauv \tauv'} \Pi_0^{\tauv'} \Pi_2^{\tauv} 
\nonumber\\
&\qquad + m_{\tauv}^* \omega v_{39,\tauv \tauv'} \Pi_0^{\tauv'} \Pi_0^{\tauv}
+ m_{\tauv}^* \omega v_{40,\tauv \tauv'} \Pi_2^{\tauv'} \Pi_0^{\tauv} \Big).
\end{align}

For the sake of comparison with the literature, we give also the expressions for the isoscalar ($I=0$) and isovector ($I=1$) RPA response functions
\begin{equation}\label{eq:full RPA response function-I0}
 \Pi_{\rm RPA}^{(SM;I=0)} = \Pi_{{\rm RPA},nn}^{(SM)} + 2 \Pi_{{\rm RPA},np}^{(SM)}  + \Pi_{{\rm RPA},pp}^{(SM)},
\end{equation}
and 
\begin{equation}\label{eq:full RPA response function-I1}
 \Pi_{\rm RPA}^{(SM;I=1)} = \Pi_{{\rm RPA},nn}^{(SM)} - 2 \Pi_{{\rm RPA},np}^{(SM)}  + \Pi_{{\rm RPA},pp}^{(SM)},
\end{equation}
respectively.

The lowest-order Landau approximation can be easily computed when we set $q=0$ and replace $p_1^2$ and $p_2^2$ by the corresponding $k_{F,\tauv}^2$ and $k_{F,\tauv'}^2$, respectively, in the expressions for the $v_{i,\tauv \tauv'}^0$. Only the coefficients $v_{1,\tauv \tauv'}$ and $v_{4,\tauv \tauv'}$ need to be solved in that case.

\section{RPA response functions in asymmetric nuclear matter at finite temperature}
\label{sec:RPA response functions in ANM}
\begin{figure}
\begin{center}
\includegraphics[scale=0.54]{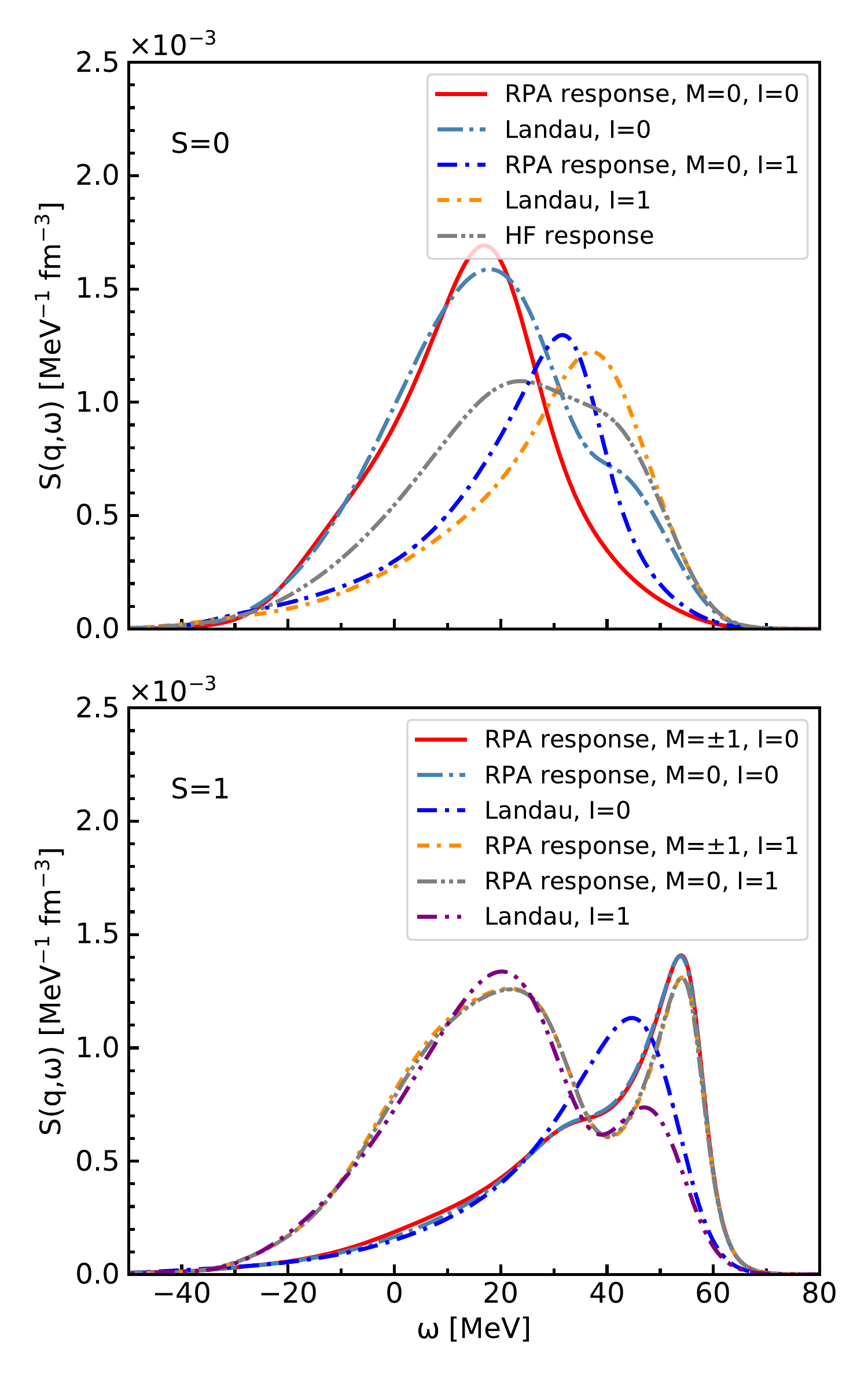} 
\caption{Isoscalar ($I=0$) and isovector ($I=1$) RPA response functions for interaction SLy5 in asymmetric nuclear matter. $\rho=0.16 $ ${\rm fm}^{-3}$, $Y_{p}=0.2$, $T=10$ MeV, $q=0.5 $ ${\rm fm}^{-1}$. Top: $S=0$ and HF response, bottom: $S=1$ response.}
\label{fig:RPA-ANM}
\end{center}
\end{figure}

\begin{figure}
\begin{center}
\includegraphics[scale=0.54]{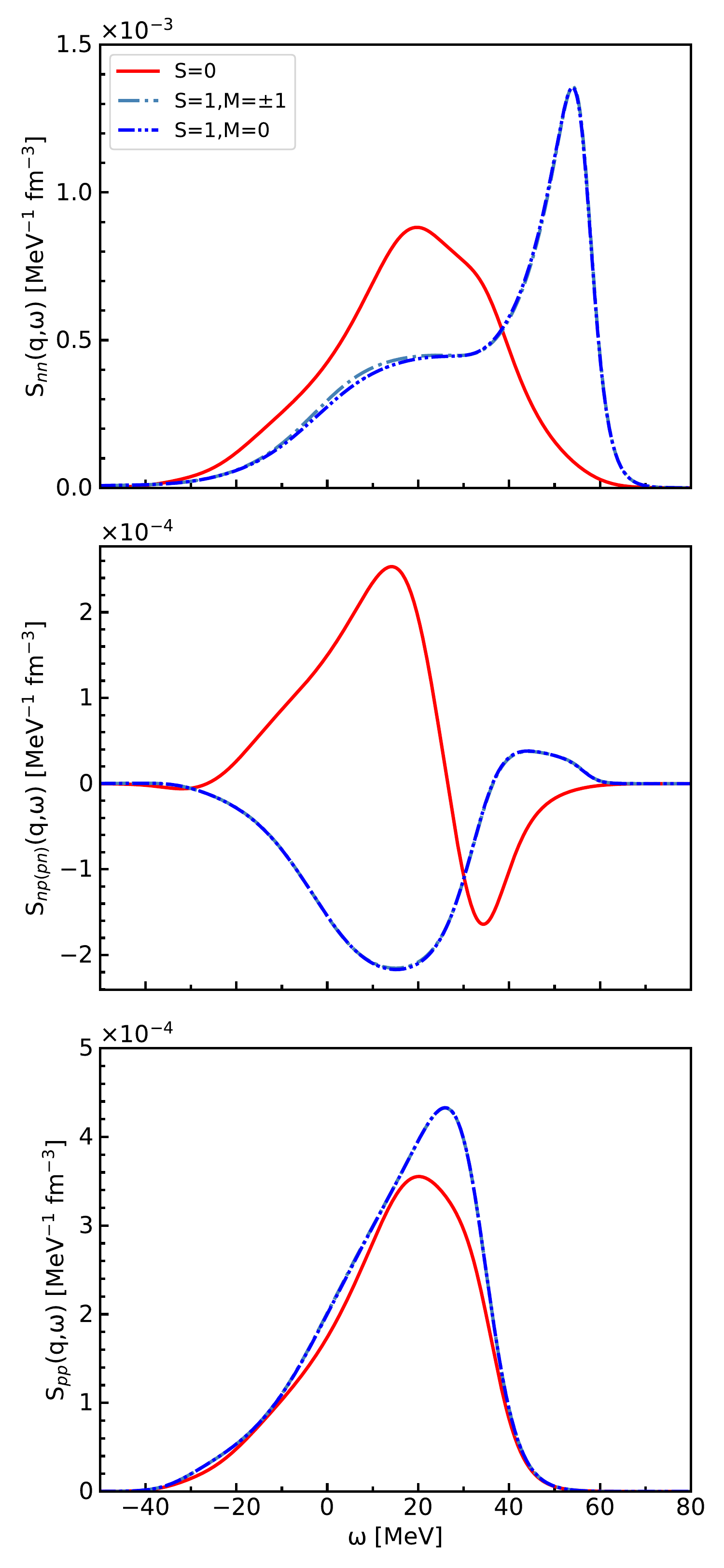} 
\caption{Dynamical structure factors $S_{nn}$, $S_{np} = S_{pn}$, and $S_{pp}$ (from top to bottom) for interaction SLy5 in asymmetric nuclear matter.  The same parameters as in Fig. \ref{fig:RPA-ANM} are chosen.}
\label{fig:Snn-Snp-Spn-Spp-ANM}
\end{center}
\end{figure}
We now present and discuss the RPA response functions in asymmetric nuclear matter at finite temperature. To check the correctness of our RPA solution, we have reproduced some of the zero-temperature results of \citep{Pastore2015}, for pure neutron matter, symmetric nuclear matter, and asymmetric nuclear matter\footnote{We checked the SLy5 results shown in Figs. 13 and 25 of \cite{Pastore2015}. In some cases, we found minor differences, but the authors of \citep{Pastore2015} confirmed that our results were correct \citep{Davesne-private}.}.
As an example for response functions at finite temperature, we show in Fig. \ref{fig:RPA-ANM} the dynamical structure factors $S(q,\omega)$ in different channels of asymmetric nuclear matter for the parametrization SLy5 of the Skyrme interaction built by Chabanat et al. \citep{Chabanat1998} (cf. Table \ref{table:parameters for skyrme interactions} which summarizes the parameters of all interactions used in this paper). For comparison, the Hartree-Fock (HF) response (i.e., $\Pi_0$) and the lowest-order Landau approximation are also shown. We directly use the proton fraction $Y_{p}$ instead of the isospin asymmetry factor defined in Refs. \citep{Davesne2014,Davesne2019} to specify the isospin asymmetry. The pure neutron matter case corresponds to $Y_{p}=0$, and the symmetric nuclear matter case corresponds to $Y_{p}=0.5$.

Fig. \ref{fig:RPA-ANM} shows the results computed for a momentum transfer $q=0.5$ fm$^{-1}$ at density $\rho = 0.16$ fm$^{-3}$ and temperature $T=10$ MeV. The proton fraction is $Y_{p}=0.2$. In the upper panel, results for channels $S=0$ and HF response are shown. Isoscalar ($I=0$) and isovector ($I=1$) responses for full RPA have both a single-broad-peak structure. In Landau approximation, the isoscalar ($I=0$) response has two bumps that stem from the neutron and proton responses, similar to the HF response and in contrast to the full RPA. For the isovector ($I=1$) responses, the difference between full RPA and Landau approximation is quite big around $\omega = 40$ MeV and the peak is shifted, but there are no additional bumps.

In the lower panel, results for channels $S=1$ are shown. The difference between $M=0$ and $M=\pm 1$ due to spin-orbit interaction is small. The full RPA responses show two bumps, the bump at high energy transfer can be considered a reminiscence of zero sound mode. Isoscalar ($S=1,I=0$) response in Landau approximation is missing this enhancement. So far, we do not see good agreement between full RPA and Landau approximation. It is also a pity that there is such a big difference between the peaks near $\omega = 55$ MeV, although the peak around $\omega = 20$ MeV of the isovector response in Landau approximation agrees quite well with that of full RPA.

In a word, the HF response and the lowest-order Landau approximation fail to reproduce the full RPA. Hence, the full RPA responses are required even though they are difficult to get and take more time to compute.

Furthermore, we present response functions as required for the neutrino scattering rates, i.e., for the different combinations $nn$, $np$, and $pp$ when the spin quantum numbers are fixed. The results are shown in Fig. \ref{fig:Snn-Snp-Spn-Spp-ANM}. The same interaction (SLy5) and parameters are adopted as before. It is obvious that there are different shapes and values between $S_{nn}$ and $S_{pp}$ since the neutron and proton densities are not equal. As mentioned in Eq. (\ref{eq:S-real}), $S_{np} $ and $S_{pn}$ are the same for the corresponding channels. As can be shown from Eqs. (\ref{eq:structure factor S-s0}) and (\ref{eq:structure factor S-s1}), $S_{nn}$ and $S_{pp}$ are positive. However, $S_{np}=S_{pn}$ can be positive or negative, although there are no negative values in the isoscalar and isovector combinations of these four dynamical structure factors (cf. Fig. \ref{fig:RPA-ANM}). This implies that $S_{nn} + S_{pp} \geq 2 S_{np(pn)}$.

\section{neutrino scattering rates}
\label{sec:neutrino scattering rates}
Neutrino-nucleon scattering is an essential part of neutrino transport in supernova simulations and proto-neutron star evolution. However, there are some problems in previous studies as mentioned in the introduction. Often the inelastic scattering reactions were neglected as in Refs. \citep{Oertel2020,Connor2015}, or nucleons were treated as an ideal gas as in Ref. \citep{Pascal2022}. Moreover, the neutrino scattering expressions given in Ref. \citep{Reddy1999} were obtained within the Landau approximation. Ref. \citep{Reddy1999} also reports an instability in Skyrme interaction SLy4 in $\beta$ equilibrated matter at $2\rho_0$. But in our examination, it turns out that there is no instability because SLy4 was fitted without $\mathbbm{J}^2$ term \citep{Chabanat1998} and $\eta_{J}$ (in the notation of \citep{Bender2003}) should be 0, thus avoiding the instability as also pointed out in Ref. \citep{Chamel2010}. The instability can be found in SLy5 which was fitted with the $\mathbbm{J}^2$ term \citep{Chabanat1998} ($\eta_{J}=1$). For this reason, we will in the rest of this paper use SLy4 instead of SLy5.

We find that the relationship between neutrino scattering rates and the scattering angle, the energy dependence of the average scattering angle, and so on, have not yet been extensively investigated in the literature. We now study neutrino scattering rates of proto-neutron star and supernova matter computed using various approximations. Especially, we will compute the inelastic neutrino scattering rates using full RPA response functions.

\subsection{Angle dependence of neutrino scattering rates in proto-neutron star and supernova matter}
\label{subsec:neutrino scattering rates evolution}
Let us start by obtaining the expression of the double-differential neutrino scattering rate with RPA response functions. According to Eq. (\ref{eq:double differential of scattering rate}), with the combination of Eqs. (\ref{eq:structure factors and correlation functions}), (\ref{eq:Pi-s0}), (\ref{eq:Pi-s1-m1}), and (\ref{eq:Pi-s1-m0}), the double-differential neutrino scattering rate can be written as 
\begin{widetext}
\begin{multline}\label{eq:scattering rate-sum}
\frac{d^2 R}{d \cos\theta d E'_{\nu}} =  \frac{G^2}{8 \pi} E_{\nu}^{\prime\,2}  \sum_{aa'=n,p}\bigg\{ (1+ \cos\theta) C_{V,a}C_{V,a'} S_{aa'}^{(S=0)}
 + 2 \left  [\frac{E_{\nu} E'_{\nu} {\rm sin}^2 \theta}{q^2} + 1- \cos\theta \right] C_{A,a} C_{A,a'} S_{aa'}^{(S=1,M=\pm1)}\\
 + \left [\frac{2(E'_{\nu} \cos\theta - E_{\nu}) (E'_{\nu} - E_{\nu} \cos\theta)}{q^2} + 1- \cos\theta \right]  C_{A,a}C_{A,a'} S_{aa'}^{(S=1,M=0)}\bigg\},
\end{multline}
\end{widetext}
where in the response functions $\omega=E_{\nu}-E'_{\nu}$ and $q=\sqrt{E_{\nu}^2 + E_{\nu}^{\prime\,2} -2 E_{\nu} E'_{\nu} \cos\theta}$. Equation (\ref{eq:scattering rate-sum}) generalizes the expression given in Eq. (71) of Ref. \citep{Pastore2015} to the case of asymmetric matter.\footnote{There is a misprint in Eq. (71) of \citep{Pastore2015}: the factor $E_\nu^2$ on the right-hand side should read $E_{\nu'}^2$ ($= E_\nu^{\prime\,2}$ in our notation).}

Since neutrinos travel with the speed of light, we will in practice present the scattering rate divided by $c$, which gives the number of collisions per traveled distance, i.e., the inverse of the mean-free path $\lambda$.

For the discussion of kinematic and RPA effects, the variables $\omega$ and $q$ are more convenient than $E'_\nu$ and $\theta$. The corresponding double-differential rate $\frac{d^2 R}{dq d\omega}$ can be computed according to the relation presented in Eq. (\ref{eq:relate-double differential}).

\begin{figure}
\begin{center}
\includegraphics[scale=0.54]{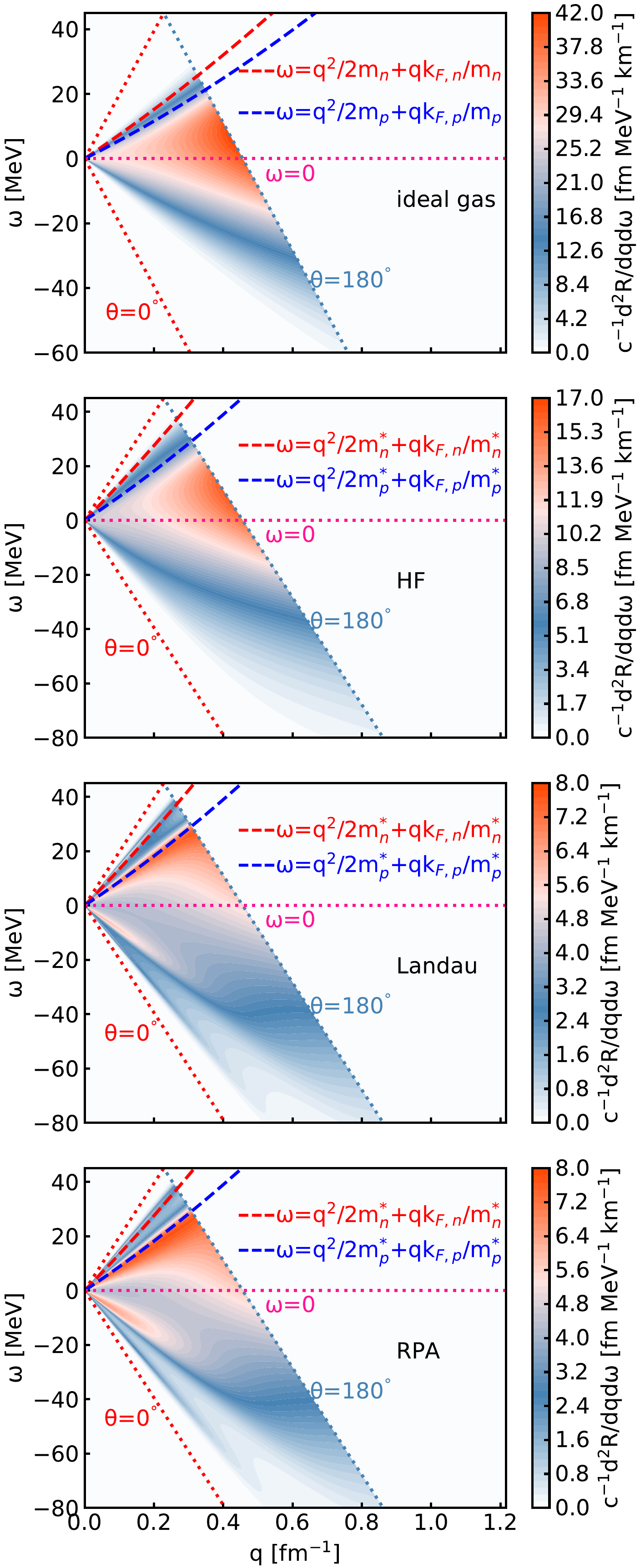} 
\caption{Double-differential neutrino scattering rates computed using four 
approximations [from top to bottom: ideal gas, Hartree-Fock (i.e., including the effective masses), Landau approximation, and full RPA] at
$T=15$ MeV, $\rho_{b}=0.25$ fm$^{-3}$, and $Y_{p}=0.296$. The initial neutrino energy
is $E_{\nu} = 3T$. The Skyrme interaction SLy4 is chosen.}
\label{fig:R-q-w}
\end{center}
\end{figure}
In the recent article \citep{Pascal2022}, the results of the simulation of proto-neutron star
evolution with improved charged-current neutrino-nucleon interactions are reported. The RG(SLy4)
EoS \citep{Gulminelli2015} constructed using Skyrme interaction SLy4 \citep{Chabanat1998} is
employed in the simulation. The baryon density and the corresponding
proton fraction for different temperatures can be found in Table 1 of \citep{Pascal2022}. We
first compute the double-differential neutrino scattering rates
$\frac{d^2 R}{dq d\omega}$ for the Skyrme interaction SLy4. We select temperature $T=15$ MeV, baryon number
density $\rho_{b}=0.25$ fm$^{-3}$, and proton fraction $Y_{p}=0.296$ from \citep{Pascal2022}. The
initial neutrino energy is fixed as $E_{\nu}=3T$. The results computed using four approximations
are shown in Fig. \ref{fig:R-q-w}. The red and steel-blue short dashed lines correspond to
scattering angles $\theta = 0^{\circ}$ and $180^{\circ}$,
respectively. The red and blue long dashed lines represent, respectively, the maximum
possible energy that can be transferred to neutrons and protons at zero temperature, i.e.,
$\omega=\frac{q^2}{2m_{a}}+\frac{qk_{F,a}}{m_a}$ ($a=n,p$) in the first panel, and $\frac{q^2}{2m_{a}^*}+\frac{qk_{F,a}}{m_a^*}$ in the other panels.
Obviously, the maximum scattering angle can be $180^{\circ}$, but the minimum scattering angle
cannot be $0^{\circ}$ except for elastic scattering ($\omega=0$). There is no contribution
from protons when the combination of energy and momentum transfers ($q,\omega$) is located in the
region between the two long-dashed lines. There is also such a region below $\omega=0$. These two
regions are symmetric with respect to $\omega\leftrightarrow-\omega$: at $\omega>0$ the neutrino excites the
nuclear matter to some excited state, while at $\omega<0$ it deexcites a thermally excited
state. Because the initial neutrino energy $E_{\nu}$ is fixed, a larger energy transfer
$\omega$ corresponds to a smaller final neutrino energy $E'_{\nu}$. We can see that scattering
is not possible for too small $E'_{\nu}$. Besides, there are some large scattering rates above the red dashed lines for Landau approximation and RPA cases, which are produced due to the zero sound
mode.

Among the approximations shown in Fig. \ref{fig:R-q-w}, the largest scattering rates are reached for the case that nucleons are treated as an ideal gas. This is consistent with the findings of Ref. \citep{Reddy1999}.

To complete this study, we also present the evolution of neutrino scattering rates as functions of the
cosine of the angle. In addition to SLy4, we use another popular Skyrme 
interaction in the field of astrophysics, BSk20 \citep{Goriely2010}. As examples, we 
compute the neutrino scattering rates at $T=3.75$ MeV, $\rho_b = 0.196$ fm$^{-3}$, and $Y_p = 0.054$ (also from Ref. \cite{Pascal2022}) in
addition to the combination $T=15$ MeV, $\rho_b = 0.25$ fm$^{-3}$, and $Y_p = 0.296$ used already in Fig. \ref{fig:R-q-w}. On the 
one hand, $T=3.75$ MeV and $T=15$ MeV represent a low temperature and an intermediate temperature, 
respectively. On the other hand, according to Table 1 of \citep{Pascal2022}, baryons consist of only free neutrons and protons and no nuclei are present under these conditions. To compare, we choose the same temperature, baryon number density, and proton 
fraction for BSk20 as for SLy4, although we do not know the exact combinations that would be obtained if the interaction BSk20 was employed in the
simulation. 

\begin{figure}
\begin{center}
\includegraphics[scale=0.54]{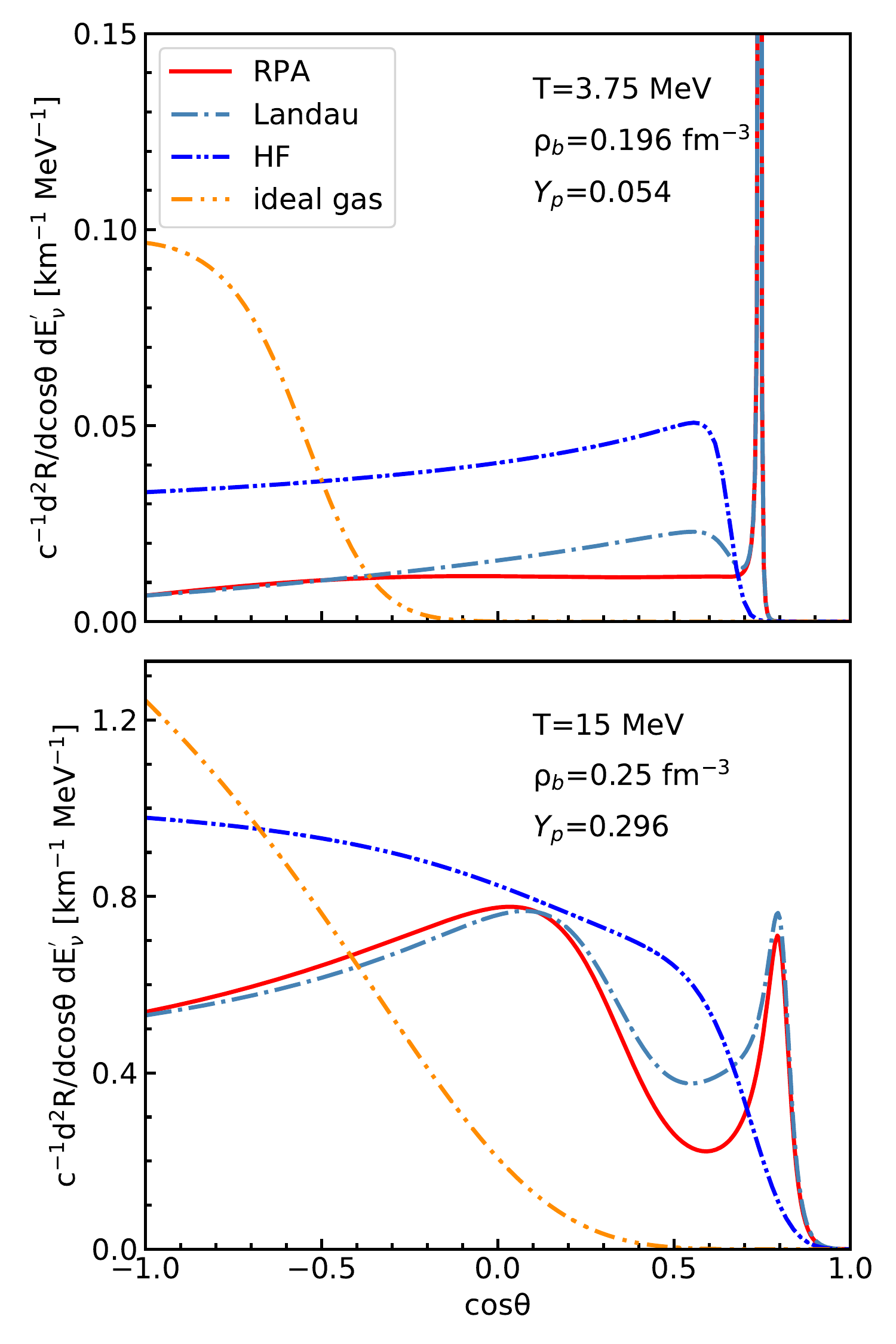} 
\caption{The neutrino scattering rates as a function of $\cos\theta$ with the initial neutrino energy $E_{\nu} = 3T$ and the final neutrino energy $E'_{\nu} = \frac{E_{\nu}}{2}$, the interaction SLy4 is chosen. Upper panel: $T = 3.75$ MeV, $\rho_b= 0.196$ fm$^{-3}$, $Y_{p}=0.054$. Lower panel: $T = 15$ MeV, $\rho_b= 0.25$ fm$^{-3}$, $Y_{p}=0.296$.}
\label{fig:neutrino scattering rates as a function of cos theta for SLy4}
\end{center}
\end{figure}
Figure \ref{fig:neutrino scattering rates as a function of cos theta for SLy4} shows the evolution of 
neutrino scattering rates as functions of the cosine of the angle for fixed initial and final neutrino 
energies, SLy4 interaction is chosen. The upper and lower panels show the results for temperature 
$T=3.75$ MeV and $T=15$ MeV, respectively. The initial and final neutrino energies are fixed to $E_{\nu} = 3T$ and $E'_{\nu} = \frac{E_{\nu}}{2}$. As we can see, the 
neutrino scattering rates decrease with the decreasing scattering angle (increasing $\cos\theta$) when the nucleons are treated as an ideal gas. The neutrino rates computed with the improved approximations, such as the HF response, the lowest-order Landau approximation, and the full RPA 
responses, have various evolution patterns. The minimum scattering angle of the ideal gas case is larger 
than that computed with the other approximations. That is because the effective mass 
reduces the minimum scattering angle. For example, at zero temperature, one has
\begin{equation}\label{eq:omega}
\omega \leq \frac{q^2}{2m_n^*} + \frac{k_{F,n} q}{m_{n}^*}
\end{equation}
(except if there is zero sound), and from this one can show the allowed values for $\cos\theta$ for fixed $E_{\nu}$, $E'_{\nu}$ ($\omega = E_{\nu}-E'_{\nu}$):
\begin{equation}\label{eq:costheta1}
 \cos\theta \leq 1-\frac{1}{2E_{\nu} E'_{\nu}} \Big[\Big(\sqrt{k_{F,n}^2+2m_{n}^* \omega} - k_{F,n}\Big)^2-\omega^2\Big].
\end{equation}
For $\omega \ll \frac{k_{F,n}^2}{2m_n^*}$, this reduces to
\begin{equation}\label{eq:costheta2}
\cos\theta \lesssim 1-\frac{\omega^2}{2E_\nu E'_\nu v_{F,n}^2}\Big(1-v_F^2-\frac{m^*_n\omega}{k_{F,n}^2}+\cdots\Big)
\end{equation}
where $v_{F,n}=\frac{k_{F,n}}{m_{n}^*}$ denotes the Fermi velocity of the neutrons. So, the maximum $\cos\theta$ for the case that neutrons have a lower effective mass is larger, giving a smaller scattering angle. The rates computed with the lowest-order Landau approximation are in qualitative agreement with those computed with the full RPA responses. They have a sharp peak near the maximum $\cos\theta$ that originates from the $S=1$ zero sound. The minimum scattering angle in the upper figure ($\rho_b = 0.196$ fm$^{-3}$) is larger than that in the lower figure ($\rho_b = 0.25$ fm$^{-3}$) for each approximation because $m^*_n$ decreases with increasing density.

\begin{figure}
\begin{center}
\includegraphics[scale=0.54]{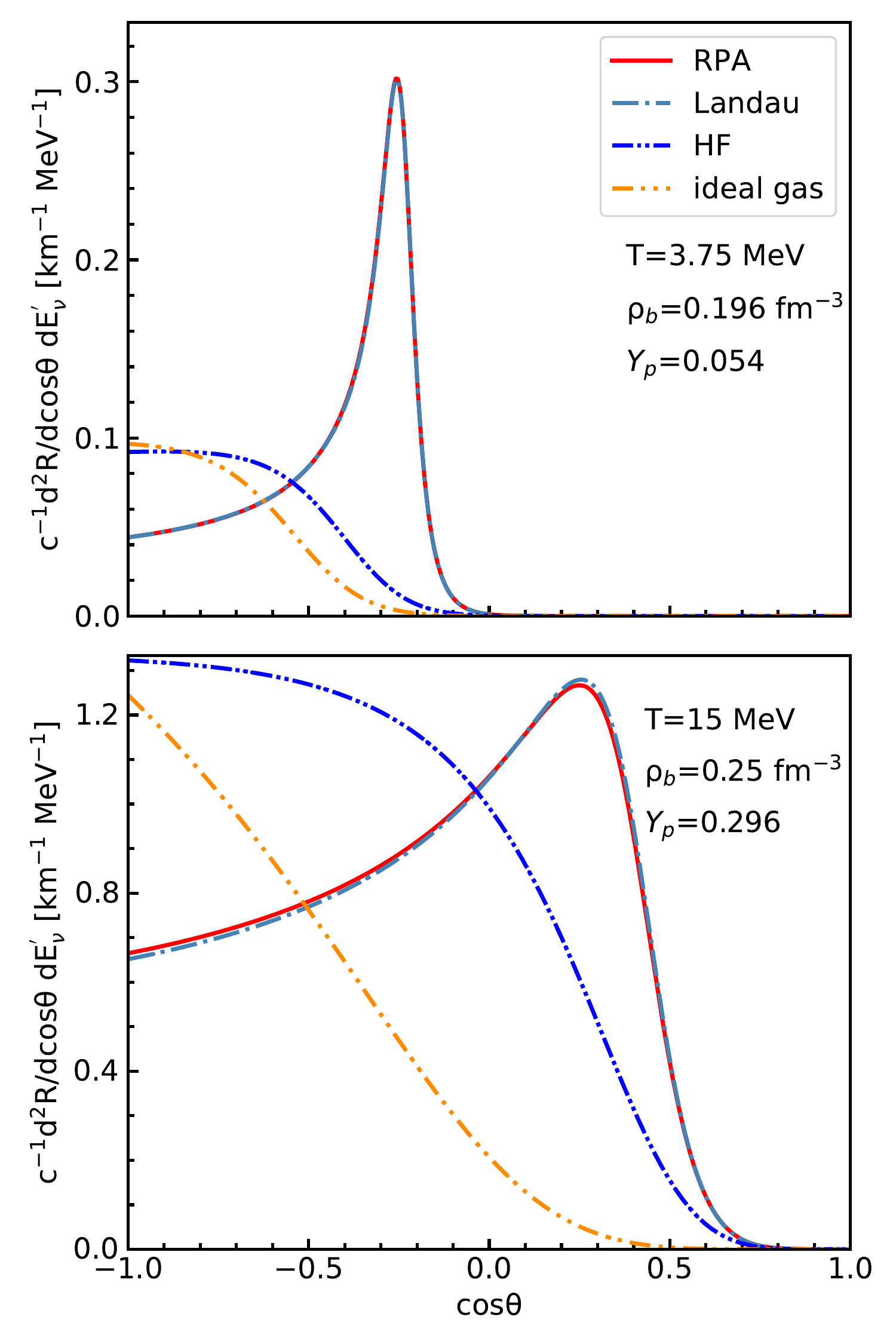} 
\caption{Similar to Fig. \ref{fig:neutrino scattering rates as a function of cos theta for SLy4} for interaction BSk20.}
\label{fig:neutrino scattering rates as a function of cos theta for BSk20}
\end{center}
\end{figure}
Figure \ref{fig:neutrino scattering rates as a function of cos theta for BSk20} is analogous to Fig. \ref{fig:neutrino scattering rates as a function of cos theta for SLy4} but for interaction BSk20. The neutrino scattering rates are decreasing with decreasing scattering angle for ideal gas cases as well as HF computations. The other curves have a broad maximum near the minimum scattering angle (maximum $\cos\theta$). Surprisingly, the results computed using the Landau approximation agree very well with those computed using the full RPA responses. Moreover, compared to Fig. \ref{fig:neutrino scattering rates as a function of cos theta for SLy4}, the minimum scattering angle is larger than in the case of SLy4 for both two temperatures because the neutron effective mass is larger with BSk20 than with SLy4.

The analysis above shows that the neutrino scattering rates in proto-neutron star and supernova matter depend sensitively on the adopted interaction and the selected approximations. In particular, the minimum scattering angle (maximum $\cos\theta$) differs for different interactions because it depends on the Fermi velocity which in turn depends on the effective mass.

\subsection{Problem of the neutron Fermi velocity}
\label{subsec:neutron Fermi velocity}

As mentioned above, the minimum scattering angle is related to the Fermi velocity. Therefore, we now study the neutron Fermi velocity. 

As an example, we take neutron-star matter under the condition of $\beta$ equilibrium at zero temperature. The fractions $Y_i$ can be obtained by combining the conditions of charge neutrality, $\rho_p = \rho_e+\rho_\mu$, and chemical equilibrium, $\mu_n - \mu_p = \mu_e = 
\mu_\mu$, where $\rho_e$, $\rho_\mu$, $\mu_e$, and $\mu_\mu$ are the electron and muon densities and chemical potentials, respectively 
\cite{Chabanat1997}. Once the neutron and proton fractions are obtained as functions of baryon number density, the corresponding neutron effective mass and Fermi momentum can be given, thus the Fermi velocity.

\begin{figure}
\begin{center}
\includegraphics[scale=0.54]{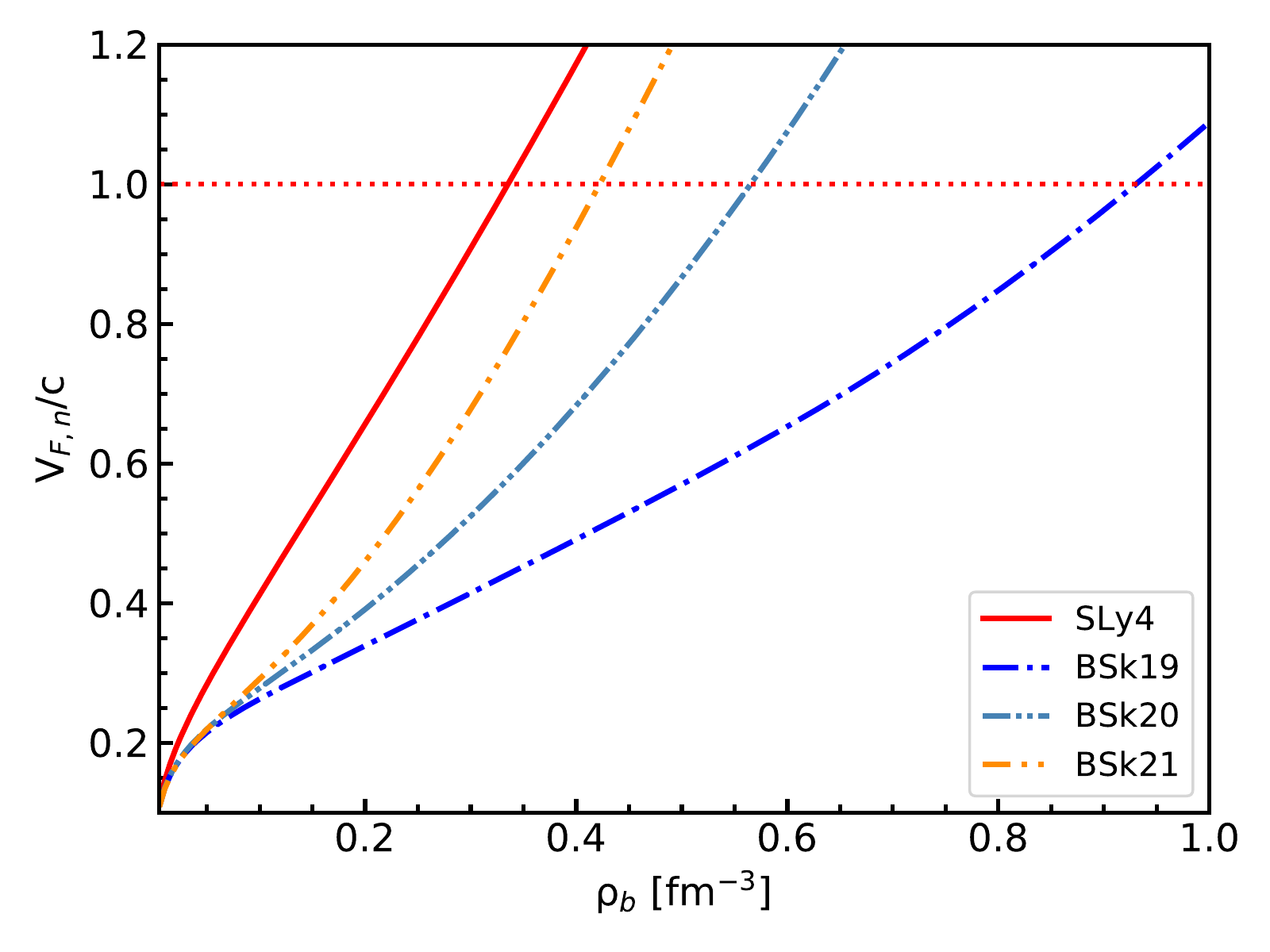} 
\caption{Fermi velocity of neutrons as a function of the density in neutron-star matter under the condition of $\beta$ equilibrium at zero temperature.}\label{fig:fermi velocity}
\end{center}
\end{figure}
Figure \ref{fig:fermi velocity} shows the Fermi velocity of neutrons as a function of the density for four Skyrme interactions, SLy4, BSk19, BSk20, and BSk21 \citep{Chabanat1998,Goriely2010}, which have been widely applied to astrophysical research. The maximum central densities of non-rotating neutron stars predicted by them are 1.21 fm$^{-3}$, 1.45 fm$^{-3}$, 0.98 fm$^{-3}$, and 0.97 fm$^{-3}$, respectively \citep{Fantina2013}. 

While one generally pays attention to the speed of sound, asking that it should be less than the speed of light \citep{Goriely2010,Chabanat1997}, Fig. \ref{fig:fermi velocity} indicates that the Fermi velocity of neutrons in neutron-star matter exceeds the speed of light at a density below the maximum central density of the neutron star predicted by these four Skyrme interactions. It means that these non-relativistic interactions should not be used at such high densities. In this respect, the interaction BSk20 seems to be better than SLy4 if we need to
choose one of them. As shown in Fig. \ref{fig:fermi velocity}, the density where the neutron Fermi velocity exceeds the speed of light is larger for BSk20 than for SLy4. Concerning BSk19, this density is even higher, but this interaction is too soft and its maximum mass of a neutron star is too low. 

An obvious solution for this problem could be to employ, at least at high densities, a relativistic theory. Another option would be to use interactions that give a larger effective mass. For instance, the so-called KIDS Skyrme-like functionals considered in Ref. \citep{Hutauruk2022} predict a neutron effective mass that is even larger than the free nucleon mass and we have checked that both neutron and proton Fermi velocities stay well below the speed of light at all relevant densities in this case.

\begin{figure*}
\begin{center}
\includegraphics[scale=0.54]{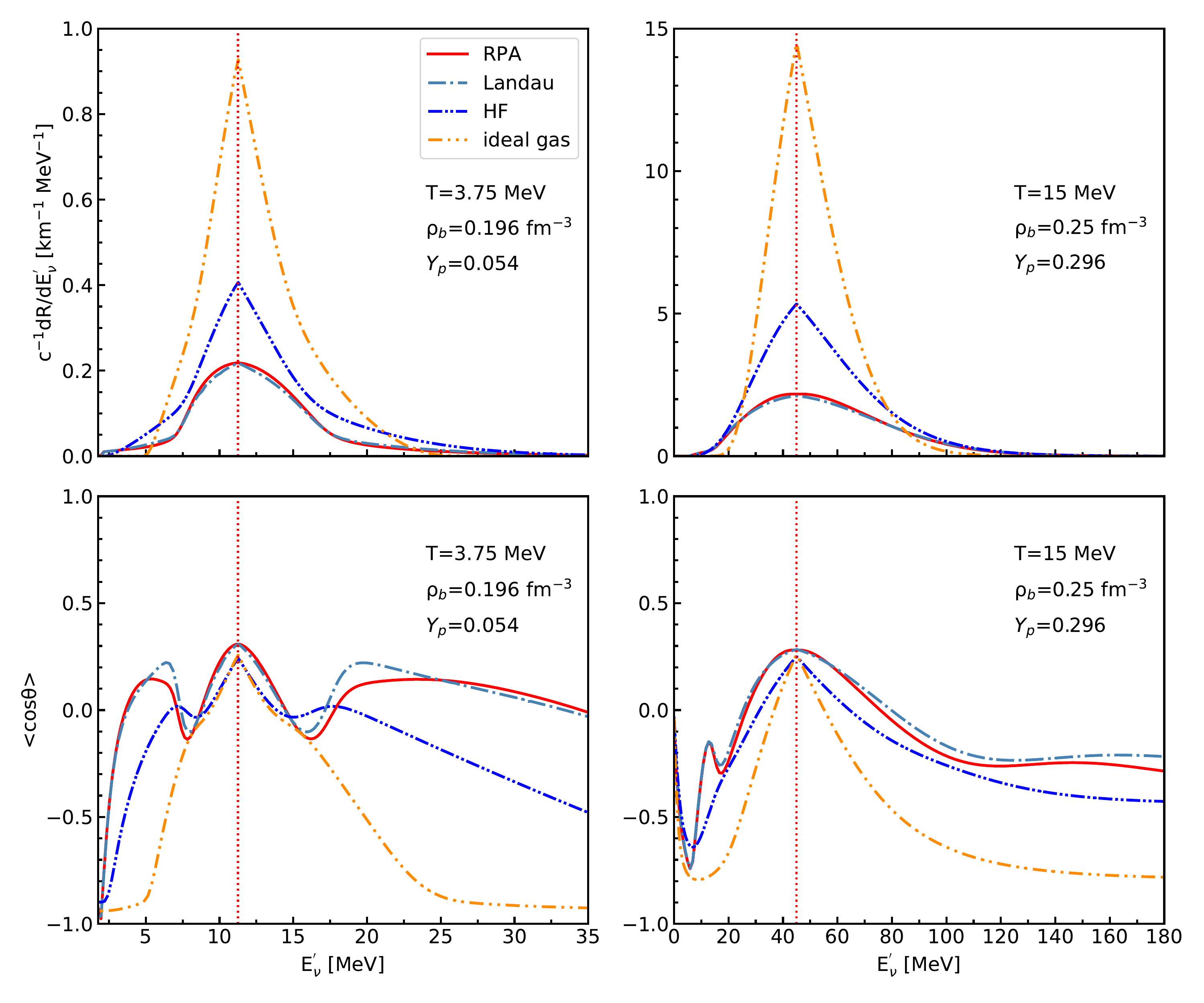} 
\caption{Differential neutrino scattering rates as a function of the final neutrino energy $E'_{\nu}$ with the fixed initial neutrino energy $E_{\nu} = 3T$ for interaction SLy4 (upper panels) and the corresponding evolution of the average cosine of the scattering angle (lower panels). The left panels are at $T=3.75$ MeV, $\rho_b = 0.196$ fm$^{-3}$, $Y_p = 0.054$, and the right panels are at $T=15$ MeV, $\rho_b = 0.25$ fm$^{-3}$, $Y_p = 0.296$.}\label{fig:neutrino scattering rates as a function of E1 for SLy4}
\end{center}
\end{figure*}

\begin{figure*}
\begin{center}
\includegraphics[scale=0.54]{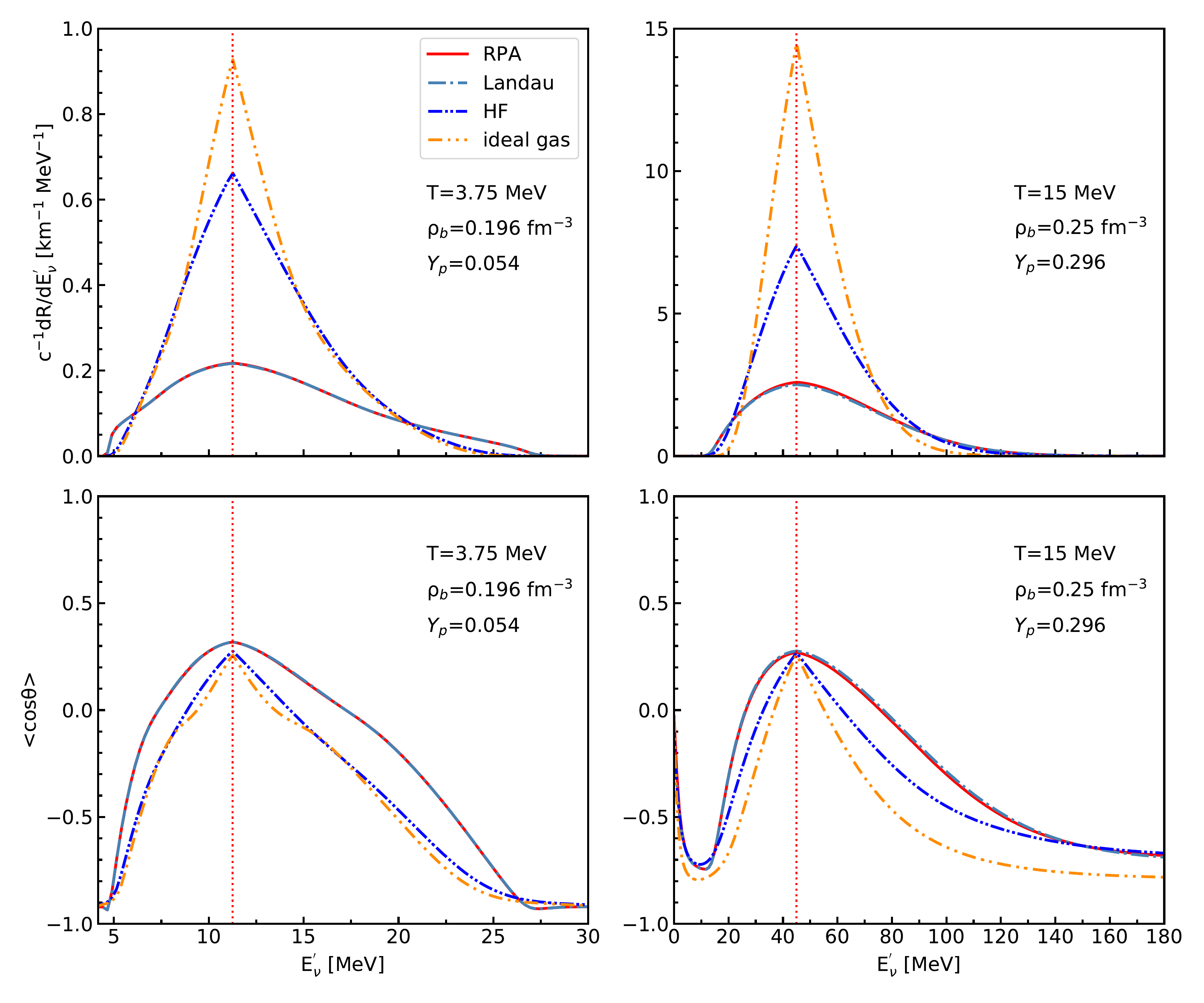} 
\caption{Similar to Fig. \ref{fig:neutrino scattering rates as a function of E1 for SLy4} for interaction BSk20.}\label{fig:neutrino scattering rates as a function of E1 for BSk20}
\end{center}
\end{figure*}

\subsection{Energy dependence of neutrino scattering rates and average 
scattering angle evolution}
\label{subsec:rates and angle}

Although there exists a problem with Skyrme interactions, the unphysical 
feature is not present at the densities and temperatures that we chose in 
Section \ref{subsec:neutrino scattering rates evolution}. Now we continue 
to study the energy dependence of neutrino scattering rates and average 
scattering angle under these conditions. 

First, we study the dependence on the final neutrino energy of the 
neutrino scattering rates and of the corresponding average cosine of the 
scattering angle. In supernova simulation codes, one sometimes expands the 
neutrino-nucleon scattering kernel in the first two terms of a Legendre 
series \citep{Bruenn1985}:
\begin{equation}\label{eq:scattering kernels}
\frac{d^2R}{dE'_\nu\,d\cos\theta}=\frac{1}{2}R_0(E_{\nu},E'_{\nu}) + 
\frac{3}{2} R_1(E_{\nu},E'_{\nu})\cos\theta,
\end{equation}
where $R_l(E_{\nu},E'_{\nu})$ ($l=0,1$) can be defined as 
\begin{equation}\label{eq:Rl}
R_l(E_{\nu},E'_{\nu}) = \int_{-1}^{1} d \cos\theta 
\frac{d^2 R}{dE'_\nu\,d\cos\theta} P_l(\cos\theta),
\end{equation}
with the Legendre polynomials $P_0(\cos\theta)=1$, 
$P_1(\cos\theta)=\cos\theta$. Then the coefficients $R_0$ and $R_1$ of 
the Legendre series are directly related to the differential neutrino 
scattering rate and average cosine of the scattering angle through the 
relations
\begin{equation}\label{eq:differential neutrino scattering rate}
\frac{dR}{dE'_{\nu}} = \int_{-1}^{1} d \cos\theta 
\frac{d^2 R}{dE'_{\nu}\,d\cos\theta} 
= R_0(E_{\nu},E'_{\nu}),
\end{equation}
and
\begin{equation}\label{eq:average scattering angle}
\langle \cos\theta \rangle = 
\frac{R_1(E_{\nu},E'_{\nu})}{R_0(E_{\nu},E'_{\nu})}.
\end{equation}

The upper panels in Fig. 
\ref{fig:neutrino scattering rates as a function of E1 for SLy4} show the 
differential neutrino scattering rates as functions of the final neutrino 
energy $E'_{\nu}$ with the fixed initial neutrino energy $E_{\nu} = 3T$ 
for interaction SLy4 at the same conditions as in Fig. 
\ref{fig:neutrino scattering rates as a function of cos theta for SLy4},
i.e.,  $T=3.75$ MeV (left panel) and $T=15$ MeV (right panel). The lower 
panels are the corresponding average cosine of the scattering angle 
$\langle \cos\theta \rangle $. The red short dashed lines represent 
$E'_{\nu}=E_{\nu}$. As noted before, in a broad range of $E'_{\nu}$, the 
neutrino scattering rates computed by treating the nucleons as an ideal 
gas are the largest, followed by those computed using the HF response. 
Those computed using the Landau approximation and the full RPA are 
smaller, although there are sharp peaks in Fig. 
\ref{fig:neutrino scattering rates as a function of cos theta for SLy4} of 
Section \ref{subsec:neutrino scattering rates evolution}. In a broad range 
of $E'_{\nu}$, the scattering angles computed by treating the nucleons as 
an ideal gas are larger (i.e., $\langle\cos\theta\rangle$ is smaller) than 
those computed using the other three approximations. The maximum 
differential neutrino scattering rates lie around $E'_{\nu} = E_{\nu}$. 
The minimum scattering angles are also found around this point. The 
differential neutrino scattering rates and average scattering angles 
computed using the Landau approximation agree well with those computed 
using the full RPA, only the scattering angles have minor differences 
between these two approximations.

Figure \ref{fig:neutrino scattering rates as a function of E1 for BSk20} is similar to Fig. \ref{fig:neutrino scattering rates as a function of E1 for SLy4} but for interaction BSk20. The $\langle\cos\theta\rangle$ curves have a simpler structure compared to Fig. \ref{fig:neutrino scattering rates as a function of E1 for SLy4}. The curves computed using the Landau approximation agree  very well with those computed using the full RPA. In a broad range of $E'_{\nu}$, we see the same qualitative behavior as in Fig. \ref{fig:neutrino scattering rates as a function of E1 for SLy4}, i.e., the rates and scattering angles are largest for the ideal gas and get reduced with the inclusion of the effective mass (HF), and even more in the RPA (both with Landau approximation or full RPA).

The neutrino scattering rates and scattering angles depend sensitively on the final neutrino energy. Neglecting the inelastic neutrino-nucleon scattering or treating the nucleons as an ideal gas is a crude approximation. The results computed using the Landau approximation and the full RPA show a good agreement. This is because, unlike in Fig. \ref{fig:RPA-ANM}, the momentum transfer $q$ in neutrino scattering is typically much smaller than the Fermi momentum. Therefore, if calculations within the full RPA are too slow, one should at least introduce the Landau approximation in the study of supernova simulation or proto-neutron star evolution. 

\begin{figure}
\begin{center}
\includegraphics[scale=0.54]{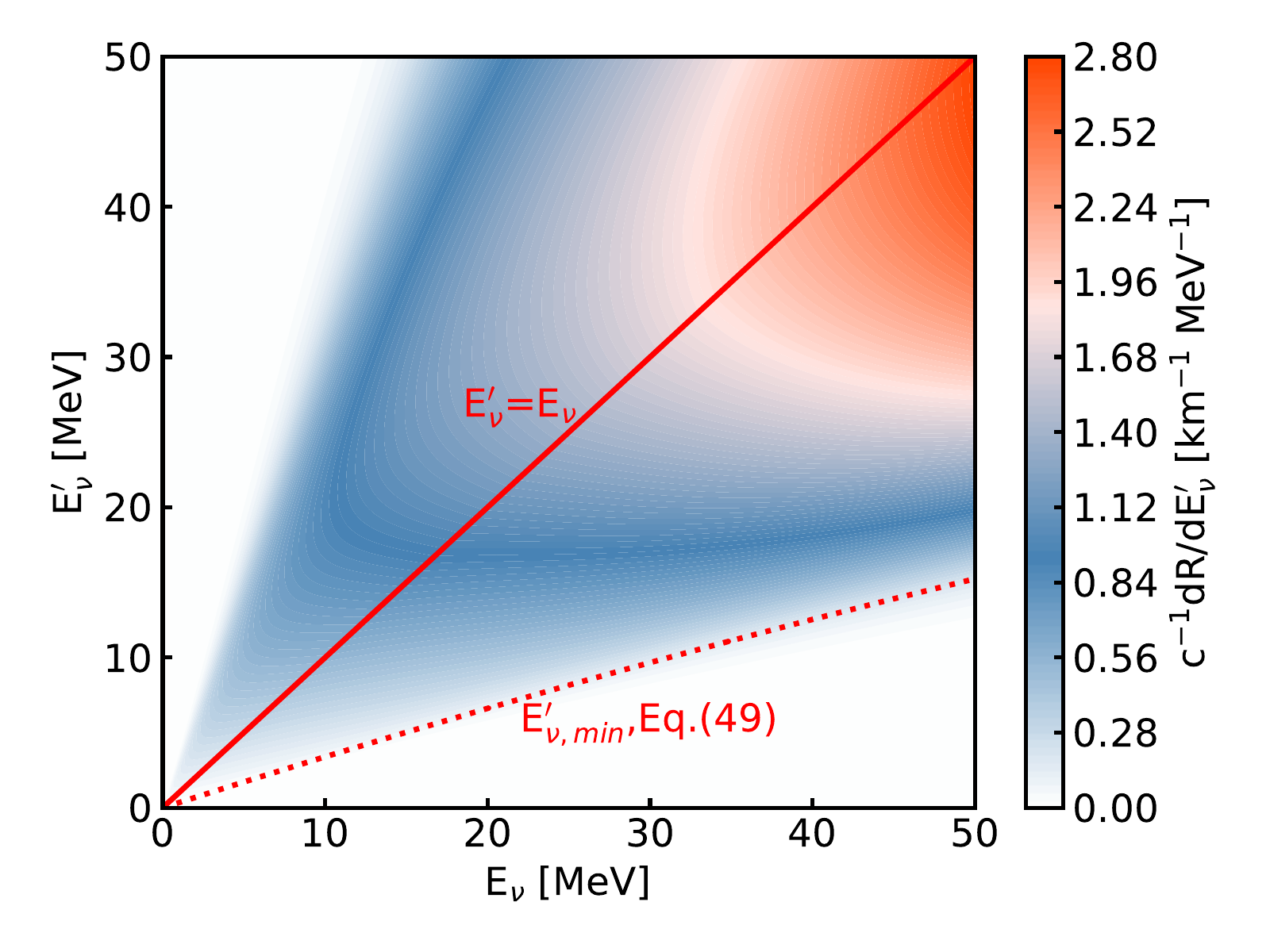}
\caption{Differential neutrino scattering rates computed using the Landau
approximation at $T=15$ MeV, $\rho_b=0.25$ fm$^{-3}$, and $Y_p=0.296$. The
Skyrme interaction BSk20 is chosen.}\label{fig:R-E-E1}
\end{center}
\end{figure}
So far, we have arbitrarily fixed the initial neutrino energy to $E_\nu = 3T$. Let us now study the dependence of the differential neutrino scattering rates on the initial and final neutrino energies. Figure \ref{fig:R-E-E1} shows the differential neutrino scattering rates computed using the Landau approximation at $T=15$ MeV, $\rho_b=0.25$ fm$^{-3}$, and $Y_p=0.296$. The Skyrme interaction BSk20 is chosen. The maximum value that the neutrino scattering rate can reach increases with increasing initial neutrino energy. Maybe, this can explain the neutrino trapping during the supernova explosion. Those neutrinos with a large initial energy scatter frequently from matter so that the neutrino mean-free path is small. They cannot escape if the neutrino mean-free path is smaller than the scale of the star. Hence, they are trapped, leading to neutrino heating. We also see that the allowed minimum and maximum values of the final neutrino energy are different for different initial neutrino energies. 

As we can see, scattering is not possible for too small $E'_{\nu}$ (as already mentioned in Section \ref{subsec:neutrino scattering rates evolution}). To get some estimate, we consider as in the discussion of Fig. \ref{fig:R-q-w} the kinematically allowed energies at $T=0$. Neglecting the zero-sound and finite-temperature contributions, $E'_{\nu,\text{min}}$ is reached at the crossing point of the red long dashed and the steel-blue short dashed lines in Fig. \ref{fig:R-q-w}, i.e., at $\cos\theta = -1$, or in other words, at $q=E_{\nu}+E'_{\nu}$:
\begin{equation}\label{eq:E'-min1}
E_{\nu}-E'_{\nu} \leq \frac{k_{F,n}(E_{\nu}+E'_{\nu})}{m_{n}^*} + \frac{(E_{\nu}+E'_{\nu})^2}{2m_{n}^*}.
\end{equation}
From this equation, the condition that $E'_{\nu}$ satisfies can be given as
\begin{equation}\label{eq:E'-min2}
E'_{\nu}\geq \sqrt{(m_n^*+k_{F,n})^2 + 4 m_n^* E_{\nu}} - m_n^* - k_{F,n} -E_{\nu}.
\end{equation}
For $E_\nu \ll m_n^*$, we get
\begin{equation}\label{eq:E'-min3}
E'_{\nu} \geq \left [ \frac{1-v_{F,n}}{1+v_{F,n}} - \frac{2E_{\nu}}{m_n^*} \frac{1}{(1+v_{F,n})^3} + \cdots \right ] E_{\nu}.
\end{equation}
Although we made some approximations when deriving this lower limit, we see in Fig. \ref{fig:R-E-E1} that, at least at $T=15$ MeV, scattering to final neutrino energies below this limit (shown as the red dotted line) is very unlikely and can probably be neglected in simulations.

Besides, detailed balance implies the relationship
\begin{equation}\label{eq:scattering rate EE'-E'E}
e^{-E_{\nu}/T}\frac{dR(E_{\nu},E'_{\nu})}{E_{\nu}^{'2} dE'_{\nu}}=e^{-E'_{\nu}/T}\frac{dR(E'_{\nu},E_{\nu})}{E_{\nu}^2 dE_{\nu}},
\end{equation}
which can be also shown from the general expressions given in Section \ref{subsec:neutrino-nucleon scattering in hot nuclear matter}. From this equation,
we can derive
\begin{equation}\label{eq:scattering rate E'E}
\frac{dR(E'_{\nu},E_{\nu})}{dE_{\nu}} = \frac{dR(E_{\nu},E'_{\nu})}{ dE'_{\nu}} \left ( \frac{E_{\nu}}{E'_{\nu}} \right )^2 e^{(E_{\nu}-E'_{\nu})/T}.
\end{equation}
Therefore, to save time, it is enough to compute only the lower half ($E'_\nu \leq E_\nu$) of the scattering kernel needed for simulations of supernova explosion and proto-neutron star evolution and we can complete it via Eq. (\ref{eq:scattering rate E'E}).

\section{Conclusion} \label{sec:conclusion}
This work aims to study energy and angle dependence of neutrino scattering rates in proto-neutron star and supernova matter within Skyrme RPA response functions. We first summarize the derivation of the scattering rate and develop the full RPA response functions in asymmetric nuclear matter at finite temperature, generalizing the method proposed in Ref. \citep{Urban2020} for the static response in pure neutron matter, as an alternative (but equivalent) to the formalism of Ref. \citep{Pastore2015}. After making a comparison between the full RPA and the other response functions, such as the HF response and the lowest-order Landau approximation, for Skyrme interaction SLy5, we conclude that in general these simpler approximations cannot reproduce the full RPA quite well.

We then compute the double-differential neutrino scattering rates using the various approximations, including ideal gas treatment of nucleons, HF response, the lowest-order Landau approximation, and the full RPA response function, for two popular Skyrme interactions, SLy4 and BSk20, at different realistic conditions of temperature, density, and proton fraction taken from the literature \citep{Pascal2022}. The results indicate that for given temperature, baryon number density, and proton fraction, the neutrino scattering rates as functions of the scattering angle depend on the adopted interaction and the selected approximation.
As it was observed in earlier studies \citep{Reddy1999}, the HF scattering rates which take into account the effective mass are sizeably smaller than those obtained for the ideal gas, and including the RPA effects reduces them further. Now, for the typical momentum transfers that occur in neutrino scattering, the Landau approximation turns out to agree well with the full RPA calculations. The minimum scattering angle is different for different interactions because it depends sensitively on the Fermi velocity which in turn depends on the effective mass.

Therefore, we study in more detail the neutron Fermi velocity. We find that in neutron star matter, it exceeds the speed of light at a density below the maximum central density of the neutron star predicted by the Skyrme interactions. While the causality of non-relativistic equations of state is routinely checked by verifying that the speed of sound is less than the speed of light, the problem of the Fermi velocity has, to our knowledge, not attracted much attention in the existing literature. To solve this problem, it may be preferable to employ a relativistic theory at high densities, or to choose interactions which do not have this problem because they predict a larger neutron effective mass.

Present neutrino transport codes cannot deal with fine details of the angle dependence such as the peak due to the zero-sound mode and the minimum scattering angle. Instead, only the rate and the corresponding average scattering angle for neutrino scattering from one energy to another can be included as described in Ref. \cite{Pascal2022}. Therefore, we study the neutrino scattering rates and average scattering angles as functions of the initial and final neutrino energies for different temperatures, densities and adopted interactions. Not only the rates, also the average scattering angles are systematically reduced as one passes from the ideal gas to the HF approximation and finally to the RPA.
This may have some consequences for supernova simulations because both the reduced rates as well as the reduced scattering angles decrease the momentum and energy that the neutrinos could transfer to the nuclear matter to revive the shock.

It would be therefore interesting to see what happens if the full rates and average scattering angles are included in the neutrino transport part of the core collapse and proto-neutron star simulation. This necessitates, however, that the results of our quite complicated calculations are parameterized or tabulated in a suitable way. This is left for future work. Furthermore, there are obviously many other open problems, such as the strong dependence on the chosen interaction, the mentioned problems of non-relativistic models at high density, and the scattering of neutrinos in matter at subsaturation densities which consists of a mixture of unbound nucleons and nuclei.

\begin{acknowledgments}
Mingya Duan is grateful for the support of China Scholarship Council (CSC No. 202006660002).
\end{acknowledgments}

\begin{widetext}
\appendix*
\section{Skyrme parameters, generalized Lindhard functions $\Pi_i$, coefficients $v_{i,\tauv \tauv'}^0$, and matrix $A_{ik}$ for the RPA response}
The parameters of the Skyrme interactions used in this article are listed in Table \ref{table:parameters for skyrme interactions}.

The generalized Lindhard functions $\Pi_i$ in this work can be written using the $\beta_i$ functions of Ref. \citep{Pastore2015}:
\begin{align}
\label{eq:compare with pastore's}
 &\Pi_0 = 2 \beta_0, 
 &&
 \Pi_2 = 2 q^2 \beta_2 + \frac{q^2}{2} \beta_0 + 2 q^2 \beta_1,
 \nonumber\\
 &
\Pi_4 = 2 q^4 \beta_5 + q^4 \beta_2 +4 q^4 \beta_4 + q^4 \beta_1 + 2 q^4 \beta_3 +\frac{q^4}{8} \beta_0,
&&\Pi_{2L} = 2 q^2 \beta_3 + \frac{q^2}{2} \beta_0 + 2 q^2 \beta_1.
\end{align}

The expressions for the required coefficients $v_{i,\tauv \tauv'}^0$ are:
\begin{align}
\label{eq:coefficients-nn}
v_{1,nn}^0 =&  (C_0^{\rho} \rho^2)'' + (C_1^{\rho})'' (\rho_{n} - \rho_{p})^2 + 4(C_1^{\rho})' (\rho_{n} - \rho_{p})  + 2 C_1^{\rho} + (C_0^{\tau} \rho)'' \tau + (C_1^{\tau})''(\rho_{n} - \rho_{p}) (\tau_{n} - \tau_{p})+2(C_1^{\tau})' (\tau_n - \tau_p) 
\notag \\
& -  \left \lbrace \frac{1}{2} \big[(C_0^{\tau} \rho)' + (C_1^{\tau})' (\rho_n - \rho_p) + C_1^{\tau} \big] + 2 \big[(C_0^{\laplacian \rho } \rho)' - C_0^{\nabla\rho} + (C_1^{\laplacian \rho})' (\rho_n - \rho_p) + C_1^{\laplacian \rho} - C_1^{\nabla\rho}\big]  \right \rbrace q^2, 
\notag \\
v_{2,nn}^0 =& (C_0^{\tau} \rho)' + (C_1^{\tau})' (\rho_n - \rho_p) + C_1^{\tau},  
\qquad 
v_{28,nn}^0 = v_{2,nn}^0, 
\qquad 
v_{3,nn}^0 = -2 (C_0^{\tau} + C_1^{\tau}), 
\notag \\
v_{4,nn}^0 =& 2(C_0^{s} + C_1^{s}) - \left [ \frac{1}{2} (C_0^{sT} + C_1^{sT}) + 2(C_0^{\laplacian s} - C_0^{\nabla\otimes s} + C_1^{\laplacian s} - C_1^{\nabla\otimes s}) \right ] q^2, 
\notag\\ 
v_{5,nn}^0 =& C_0^{sT} + C_1^{sT}, 
\qquad 
v_{6,nn}^0 = -2v_{5,nn}^0, 
\qquad
v_{8,nn}^0 = C_0^{\nabla J} +  C_1^{\nabla J}, 
\qquad v_{9,nn}^0 = v_{8,nn}^0.
\end{align}

\begin{table*}
\caption{Parameters for Skyrme interactions SLy4, SLy5, BSk19, BSk20, and BSk21 \citep{Chabanat1998,Goriely2010}.} 
\label{table:parameters for skyrme interactions}
\begin{ruledtabular}
\begin{tabular}[b]{lccccc}

& SLy4 & SLy5 & BSk19 & BSk20 & BSk21\\
\hline
$t_0$ (MeV fm$^3$) & -2488.91 & -2484.88 & -4115.21 & -4056.04 & -3961.39 \\ 
$t_1$ (MeV fm$^{5}$) & 486.82 & 483.13 & 403.072 & 438.219 & 396.131 \\
$t_2$ (MeV fm$^{5}$) &  -546.39 & -549.40 & 0 & 0 & 0\\
$t_3$ (MeV fm$^{3+3\alpha}$) & 13777.0 & 13763.0 & 23670.4 & 23256.6 & 22588.2 \\
$t_4$ (MeV fm$^{5+3\beta}$) & 0 & 0 & -60.0 & -100.000 & -100.000\\
$t_5$ (MeV fm$^{5+3\gamma}$) & 0 & 0 & -90.0 & -120.000 & -150.000 \\
$x_0$ & 0.834 & 0.778 & 0.398848 & 0.569613 & 0.885231  \\
$x_1$ & -0.344 & -0.328 & -0.137960 & -0.392047 & 0.0648452\\
$x_2$ & -1.000 & -1.000 & 0 & 0 & 0\\
$x_3$ & 1.354 & 1.267 & 0.375201 & 0.614276 & 1.03928 \\
$x_4$ & 0 & 0 & -6.0 & -3.00000 & 2.00000 \\
$x_5$ & 0 & 0 & -13.0 & -11.0000 & -11.0000\\
$t_2 x_2$ (MeV fm$^{5}$) & 0 & 0 & -1055.55 & -1147.64 & -1390.38 \\
$W_0$ (MeV fm$^{5}$) & 123.0 & 126.0 & 110.802 & 110.228 & 109.622 \\
$\alpha$ & 1/6 & 1/6 & 1/12 & 1/12 & 1/12 \\
$\beta$ & 0 & 0 & 1/3 & 1/6 & 1/2 \\
$\gamma$ & 0 & 0 & 1/12 & 1/12 & 1/12\\
$\eta_{J}$ & 0 & 1 & 0 & 0 & 0\\
\end{tabular}
\end{ruledtabular}
\end{table*}
\begin{align}
\label{eq:coefficients-pp}
v_{1,pp}^0 =&  (C_0^{\rho} \rho^2)'' + (C_1^{\rho})'' (\rho_{n} - \rho_{p})^2 - 4(C_1^{\rho})' (\rho_{n} - \rho_{p})  + 2 C_1^{\rho} + (C_0^{\tau} \rho)'' \tau + (C_1^{\tau})''(\rho_{n} - \rho_{p}) (\tau_{n} - \tau_{p}) -2(C_1^{\tau})' (\tau_n - \tau_p)
\notag \\
& -  \left \lbrace \frac{1}{2} \big[(C_0^{\tau} \rho)' - (C_1^{\tau})' (\rho_n - \rho_p) + C_1^{\tau} \big] + 2 \big[(C_0^{\laplacian \rho } \rho)' - C_0^{\nabla\rho} - (C_1^{\laplacian \rho})' (\rho_n - \rho_p) + C_1^{\laplacian \rho} - C_1^{\nabla\rho} \big]  \right \rbrace q^2, 
\notag \\
v_{2,pp}^0 =& (C_0^{\tau} \rho)' - (C_1^{\tau})' (\rho_n - \rho_p) + C_1^{\tau},  
\qquad 
v_{28,pp}^0 = v_{2,pp}^0, 
\qquad 
v_{3,pp}^0 = v_{3,nn}^0, 
\qquad
v_{4,pp}^0 = v_{4,nn}^0, 
\notag\\
v_{5,pp}^0 =& v_{5,nn}^0, 
\qquad
v_{6,pp}^0 = v_{6,nn}^0, 
\qquad
v_{8,pp}^0 = v_{8,nn}^0, 
\qquad 
v_{9,pp}^0 = v_{9,nn}^0. 
\end{align}

\begin{align}
\label{eq:coefficients-np}
v_{1,np}^0 =&  (C_0^{\rho} \rho^2)'' + (C_1^{\rho})'' (\rho_{n} - \rho_{p})^2 - 2 C_1^{\rho} + (C_0^{\tau} \rho)'' \tau + (C_1^{\tau})''(\rho_{n} - \rho_{p}) (\tau_{n} - \tau_{p}) 
\notag \\
& - \left \lbrace \frac{1}{2} \big[(C_0^{\tau} \rho)' - C_1^{\tau} \big] + 2 \big[(C_0^{\laplacian \rho } \rho)'  - C_0^{\nabla\rho} - C_1^{\laplacian \rho} + C_1^{\nabla\rho}\big]  \right \rbrace q^2, 
\notag\\
v_{2,np}^0 =& (C_0^{\tau} \rho)' - (C_1^{\tau})' (\rho_n - \rho_p) - C_1^{\tau},  
\qquad
v_{28,np}^0 = (C_0^{\tau} \rho)' + (C_1^{\tau})' (\rho_n - \rho_p) - C_1^{\tau}, 
\qquad 
v_{3,np}^0 = -2 (C_0^{\tau} - C_1^{\tau}),  
\notag \\
v_{4,np}^0 =& 2(C_0^{s} - C_1^{s}) - \left[ \frac{1}{2} (C_0^{sT} - C_1^{sT}) + 2(C_0^{\laplacian s} - C_0^{\nabla\otimes s} - C_1^{\laplacian s} + C_1^{\nabla\otimes s}) \right] q^2, 
\notag\\
v_{5,np}^0 =& C_0^{sT} - C_1^{sT}, 
\qquad 
v_{6,np}^0 = -2v_{5,np}^0, 
\qquad
v_{8,np}^0 = C_0^{\nabla J} -  C_1^{ \nabla J}, 
\qquad 
v_{9,np}^0 = v_{8,np}^0. 
\end{align}

\begin{align}
\label{eq:coefficients-pn}
&v_{1,pn}^0 = v_{1,np}^0, 
&&v_{2,pn}^0 = v_{28,np}^0, 
&&v_{28,pn}^0 = v_{2,np}^0, 
&&v_{3,pn}^0 = v_{3,np}^0, 
&&v_{4,pn}^0 = v_{4,np}^0, 
\notag\\
&v_{5,pn}^0 = v_{5,np}^0,  
&&v_{6,pn}^0 = v_{6,np}^0,  
&&v_{8,pn}^0 = v_{8,np}^0, 
&&v_{9,pn}^0 = v_{9,np}^0. 
\end{align}
For standard Skyrme forces, the expressions of the $C$ coefficients in terms of the Skyrme parameters are given in Appendix A of \citep{Bender2003}. There, integration by parts was used to rewrite the $(\nablav\rho)^2$ and $(\nablav\otimes \sv)^2$ terms as $-\rho\laplacian\rho$ and $-\sv\cdot\laplacian \sv$. But if one keeps these terms in their original form, the corresponding coefficients read
\begin{align*}
C_0^{\laplacian \rho} =& -\frac{3}{32} t_1, &
C_0^{\nabla\rho} =& \frac{1}{64}[3t_1-t_2(5+4x_2)], \nonumber\\
C_1^{\laplacian \rho} =& \frac{1}{32} t_1(1+2x_1), &
C_1^{\nabla\rho} =& -\frac{1}{64}[t_1(1+2x_1)+t_2(1+2x_2)], \nonumber\\
C_0^{\laplacian s} =& \frac{1}{32} t_1(1-2x_1), &
C_0^{\nabla\otimes s} =& -\frac{1}{64}[t_1(1-2x_1)+t_2(1+2x_2)], \nonumber\\
C_1^{\laplacian s} =& \frac{1}{32} t_1, &
C_1^{\nabla\otimes s} =& -\frac{1}{64}(t_1+t_2).\nonumber
\end{align*}
These expressions, and the remaining $C$ coefficients given in Appendix A of \cite{Bender2003}, can easily be generalized to the case of BSk19, BSk20, and BSk21 by making the following substitutions \citep{Chamel2009}:
\begin{align}\label{eq:C coefficients-substitutions}
& t_1 \rightarrow t_1+t_4 \rho^{\beta},
&& t_1x_1 \rightarrow t_1x_1 + t_4x_4 \rho^{\beta},
&& t_2 \rightarrow t_2+t_5\rho^{\gamma},
&& t_2x_2 \rightarrow t_2x_2+t_5x_5 \rho^{\gamma}.
\end{align}

The non-vanishing matrix elements $A_{ik}$ are given below. For brevity, the isospin indices are not written. To get $A_{ik,\tauv \tauv'}$, one has to replace in the following expressions $m^* \rightarrow m_{\tauv'}^*$, $v_n^0\to v_{n,\tauv \tauv'}^0$, and $\Pi_n\to \Pi_{n}^{\tauv'}$:
\begin{align}
\label{eq:matrix elements of A}
& A_{1,1} = v_1^0 \Pi_0 + v_2^0 \Pi_2
&& A_{1,28} = v_1^0 \Pi_2 + v_2^0 \Pi_4  
&& A_{1,30} =2 v_8^0 q^2  \Pi_{2T} 
&& A_{1,35}= m^* \omega(v_1^0 \Pi_0 + v_2^0 \Pi_2)
\notag \\
& A_{2,2} = v_1^0 \Pi_0 + v_2^0 \Pi_2 
&& A_{2,10}= v_1^0 \Pi_2 + v_2^0 \Pi_4 
&& A_{2,16}= 2   v_8^0 q^2 \Pi_{2T}
&& A_{2,36} = m^*\omega (v_1^0 \Pi_0 + v_2^0 \Pi_2)
\notag \\
& A_{3,3} = v_3^0 \Pi_{2T}
&& A_{3,9}= v_9^0  q^2 \Pi_0 
&& A_{3,34}=v_9^0  q^2  \Pi_2 
&& A_{3,42}= m^* \omega   v_9^0 q^2 \Pi_0 
\notag \\
& A_{4,4}=v_4^0 \Pi_0 + v_5^0 \Pi_2
&& A_{4,29}= v_4^0 \Pi_2 + v_5^0 \Pi_4
&& A_{4,31}= v_9^0  q^2 \Pi_{2T}
&& A_{4,37}= m^* \omega (v_4^0 \Pi_0 + v_5^0 \Pi_2)
\notag \\
& A_{5,5}=v_4^0 \Pi_0 + v_5^0 \Pi_2
&& A_{5,12}=v_4^0 \Pi_2 + v_5^0 \Pi_4 
&& A_{5,17}=  v_9^0 q^2 \Pi_{2T}
&& A_{5,38} = m^* \omega (v_4^0 \Pi_0 + v_5^0 \Pi_2)
\notag \\
& A_{6,6}=v_6^0 \Pi_{2T}  
&& A_{7,7}=v_4^0 \Pi_0 + v_5^0 \Pi_2
&& A_{7,31}=-v_9^0 \Pi_{2T}
&& A_{7,32}=v_4^0 \Pi_2 + v_5^0 \Pi_4
\notag \\
&A_{7,39}=m^*\omega (v_4^0 \Pi_0 + v_5^0 \Pi_2)
&& A_{8,6}=v_8^0 \Pi_{2T}
&& A_{8,8}= v_1^0 \Pi_0 + v_2^0 \Pi_2
&& A_{8,18}=2  v_8^0 q^2 \Pi_{2T}
\notag \\
& A_{8,33}=v_1^0 \Pi_2 + v_2^0 \Pi_4
&& A_{8,41}=m^*\omega(v_1^0 \Pi_0 + v_2^0 \Pi_2)
&& A_{9,3}= v_9^0 \Pi_{2T}
&& A_{9,9}= v_4^0 \Pi_0 + v_5^0 \Pi_2
\notag \\
& A_{9,34}= v_4^0 \Pi_2 + v_5^0 \Pi_4
&& A_{9,42}= m^* \omega(v_4^0 \Pi_0 + v_5^0 \Pi_2)
&& A_{10,2}= v_{28}^0 \Pi_0
&& A_{10,10}= v_{28}^0 \Pi_2
\notag \\
& A_{10,36}=m^* \omega v_{28}^0 \Pi_0
&& A_{11,3}=\frac{v_3^0 (\Pi_{2L} - \Pi_{2T})}{q^2}
&& A_{11,9}= -v_9^0 \Pi_0
&& A_{11,11}=v_3^0 \Pi_{2L}
\notag \\
& A_{11,19}=\frac{m^* \omega}{q^2} v_3^0 \Pi_0
&& A_{11,20}=\frac{m^* \omega}{q^2} v_3^0 \Pi_2 
&& A_{11,34}=-v_9^0 \Pi_2
&&  A_{11,42}=-m^* \omega v_9^0 \Pi_0
\notag \\
&A_{12,5}=v_5^0 \Pi_0
&& A_{12,12}=v_5^0 \Pi_2
&& A_{12,38}=m^* \omega v_5^0 \Pi_0
&& A_{13,6}= \frac{v_6^0 (\Pi_{2L} - \Pi_{2T})}{q^2}
\notag \\
& A_{13,13}=v_6^0 \Pi_{2L}
&& A_{13,21}=\frac{m^* \omega}{q^2} v_6^0 \Pi_0
&& A_{13,22}=\frac{m^* \omega}{q^2} v_6^0 \Pi_2
&& A_{14,14} = v_4^0 \Pi_0 + v_5^0 \Pi_2
\notag \\
& A_{14,15}= v_4^0 \Pi_2 + v_5^0 \Pi_4
&& A_{14,17}=-v_9^0 \Pi_{2T}
&& A_{14,40}=m^* \omega (v_4^0 \Pi_0 + v_5^0 \Pi_2)
&& A_{15,14}=v_5^0 \Pi_0
\notag \\
& A_{15,15}=v_5^0 \Pi_2
&& A_{15,40}=m^* \omega v_5^0 \Pi_0
&& A_{16,2}=v_8^0 \Pi_0
&& A_{16,10}= v_8^0 \Pi_2
\notag \\
& A_{16,16}=v_6^0 \Pi_{2T}
&& A_{16,36}=m^* \omega v_8^0 \Pi_0
&& A_{17,5}=v_9^0 \Pi_0
&& A_{17,12}=v_9^0 \Pi_2
\notag \\
& A_{17,17}= v_3^0 \Pi_{2T}
&& A_{17,38}=m^* \omega v_9^0 \Pi_0
&& A_{18,8}= v_8^0 \Pi_0
&& A_{18,18}=v_6^0 \Pi_{2T} 
\notag \\
& A_{18,33}= v_8^0 \Pi_2
&& A_{18,41}=m^* \omega v_8^0 \Pi_0
&& A_{19,3}=\frac{m^* \omega}{q^2} (v_1^0 \Pi_0 + v_2^0 \Pi_2)
&& A_{19,11}=m^* \omega (v_1^0 \Pi_0 + v_2^0 \Pi_2)
\notag \\
& A_{19,19}=v_1^0 \Pi_0 + v_2^0 \Pi_2
&& A_{19,20}=v_1^0 \Pi_2 + v_2^0 \Pi_4 
&& A_{19,26}=2 v_8^0 q^2 \Pi_{2T}
&& A_{20,3}=\frac{m^* \omega}{q^2} v_{28}^0 \Pi_0 
\notag \\
& A_{20,11}=m^* \omega v_{28}^0 \Pi_0
&& A_{20,19}=v_{28}^0 \Pi_0
&& A_{20,20}=v_{28}^0 \Pi_2 
&& A_{21,6}=\frac{m^* \omega}{q^2} (v_4^0 \Pi_0 + v_5^0 \Pi_2)
\notag \\
& A_{21,13}=m^* \omega (v_4^0 \Pi_0 + v_5^0 \Pi_2)
&& A_{21,21}=v_4^0 \Pi_0 + v_5^0 \Pi_2 
&& A_{21,22}=v_4^0 \Pi_2 + v_5^0 \Pi_4 
&& A_{21,27}=  v_9^0 q^2 \Pi_{2T}
\notag \\
& A_{22,6}=\frac{m^* \omega}{q^2} v_5^0 \Pi_0
&& A_{22,13}=m^* \omega v_5^0 \Pi_0
&& A_{22,21}=v_5^0 \Pi_0
&& A_{22,22}=v_5^0 \Pi_2  
\notag \\
& A_{23,23}=v_4^0 \Pi_0 + v_5^0 \Pi_2
&& A_{23,24}=v_4^0 \Pi_2 + v_5^0 \Pi_4 
&& A_{23,25}=m^* \omega (v_4^0 \Pi_0 + v_5^0 \Pi_2)
&& A_{23,27}=-v_9^0 \Pi_{2T}
\notag \\
& A_{24,23}=v_5^0 \Pi_0
&& A_{24,24}= v_5^0 \Pi_2 
&& A_{24,25}=m^* \omega v_5^0 \Pi_0 
&& A_{25,23}=\frac{m^* \omega}{q^2} v_6^0 \Pi_0
\notag \\
& A_{25,24}=\frac{m^* \omega}{q^2} v_6^0 \Pi_2 
&& A_{25,25}=v_6^0 \Pi_{2L},  
&& A_{26,3}=\frac{m^* \omega}{q^2} v_8^0 \Pi_0 
&&  A_{26,11}=m^* \omega v_8^0 \Pi_0
\notag \\
& A_{26,19}=v_8^0 \Pi_0
&& A_{26,20}=v_8^0 \Pi_2 
&& A_{26,26}=v_6^0 \Pi_{2T} 
&& A_{27,6}=\frac{m^* \omega}{q^2} v_9^0 \Pi_0 
\notag \\
& A_{27,13}=m^* \omega v_9^0 \Pi_0
&& A_{27,21}=v_9^0 \Pi_0
&& A_{27,22}=v_9^0 \Pi_2 
&& A_{27,27}=v_3^0 \Pi_{2T} 
\notag \\
& A_{28,1}=v_{28}^0 \Pi_0
&& A_{28,28}=v_{28}^0 \Pi_2 
&& A_{28,35}=m^* \omega v_{28}^0 \Pi_0
&& A_{29,4}=v_5^0 \Pi_0
\notag \\
& A_{29,29}=v_5^0 \Pi_2
&& A_{29,37}=m^* \omega v_5^0 \Pi_0
&& A_{30,1}=v_8^0 \Pi_0
&& A_{30,28}=v_8^0 \Pi_2
\notag \\
& A_{30,30}=v_6^0 \Pi_{2T}
&& A_{30,35}=m^* \omega v_8^0 \Pi_0
&& A_{31,4}=v_9^0 \Pi_0
&& A_{31,29}=v_9^0 \Pi_2
\notag \\
& A_{31,31}=v_3^0 \Pi_{2T}
&& A_{31,37}=m^* \omega v_9^0 \Pi_0
&& A_{32,7}=v_5^0 \Pi_0
&& A_{32,32}=v_5^0 \Pi_2
\notag \\
& A_{32,39}=m^* \omega v_5^0 \Pi_0
&& A_{33,8}=v_{28}^0 \Pi_0
&& A_{33,33}=v_{28}^0 \Pi_2
&& A_{33,41}=m^* \omega v_{28}^0 \Pi_0
\notag \\
& A_{34,9}=v_5^0 \Pi_0
&& A_{34,34}=v_5^0 \Pi_2
&& A_{34,42}=m^* \omega v_5^0 \Pi_0
&& A_{35,1}=\frac{m^* \omega}{q^2} v_3^0 \Pi_0
\notag \\
& A_{35,28}=\frac{m^* \omega}{q^2} v_3^0 \Pi_2
&& A_{35,35}=v_3^0 \Pi_{2L}
&& A_{36,2}=\frac{m^* \omega}{q^2} v_3^0 \Pi_0
&& A_{36,10}=\frac{m^* \omega}{q^2} v_3^0 \Pi_2
\notag \\
& A_{36,36}=v_3^0 \Pi_{2L}
&& A_{37,4}=\frac{m^* \omega}{q^2} v_6^0 \Pi_0
&& A_{37,29}=\frac{m^* \omega}{q^2} v_6^0 \Pi_2
&& A_{37,37}=v_6^0 \Pi_{2L}
\notag \\
& A_{38,5}=\frac{m^* \omega}{q^2} v_6^0 \Pi_0
&& A_{38,12}=\frac{m^* \omega}{q^2} v_6^0 \Pi_2
&& A_{38,38}=v_6^0 \Pi_{2L}
&& A_{39,7}=\frac{m^* \omega}{q^2} v_6^0 \Pi_0
\notag \\
& A_{39,32}=\frac{m^* \omega}{q^2} v_6^0 \Pi_2
&& A_{39,39}=v_6^0 \Pi_{2L}
&& A_{40,14}=\frac{m^* \omega}{q^2} v_6^0 \Pi_0
&& A_{40,15}=\frac{m^* \omega}{q^2} v_6^0 \Pi_2
\notag \\
& A_{40,40}=v_6^0 \Pi_{2L}
&& A_{41,8}=\frac{m^* \omega}{q^2} v_3^0 \Pi_0
&& A_{41,33}=\frac{m^* \omega}{q^2} v_3^0 \Pi_2
&& A_{41,41}=v_3^0 \Pi_{2L}
\notag \\
& A_{42,9}=\frac{m^* \omega}{q^2} v_6^0 \Pi_0
&& A_{42,34}=\frac{m^* \omega}{q^2} v_6^0 \Pi_2 
&& A_{42,42}=v_6^0 \Pi_{2L}
\end{align}
\end{widetext}


\bibliography{references}

\end{document}